\documentstyle[11pt,fleqn]{article}
\newcommand{\N}{\mbox{$I\!\!N$}}             %%% natural numbers
\newcommand{\R}{\mbox{$I\!\!R$}}             %%% real numbers
\newcommand{\Z}{\mbox{$Z\!\!\!Z$}}           %%% integers numbers
\newcommand{\C}{\mbox{$I\!\!\!\!C$}}         %%% complex numbers

\begin{document}

%%%%%%%%%%%%%%  Preprint number %%%%%%%%%%%%%%%%%%%%%%%%%%

\hfill{\sl preprint -  }
\par
\bigskip
\par
\rm

%%%%%%%%%%%%%   Title %%%%%%%%%%%%%%%%%%%%%%%%%%

\par
\bigskip
\LARGE
\noindent
{\bf One-loop stress-tensor renormalization in curved background:
the relation between $\zeta$-function and point-splitting approaches, and  
an improved point-splitting procedure.}
\par
\bigskip
\par
\rm
\normalsize

%%%%%%%%%%%%%%%%%%%%%%%%%%%%%%%%%%%%%%%%%%%%%

%%%%%%%%%%%% Author %%%%%%%%%%%%%%%%%%%%%%%%%%%

\large
\noindent {\bf Valter Moretti}

\large
\smallskip

\noindent 
Department of Mathematics, Trento University, and 
Istituto Nazionale di Fisica Nucleare,\\
Gruppo Collegato di Trento,
I-38050 Povo (TN), Italy.\\
E-mail: moretti@science.unitn.it

\large
\smallskip

\rm\normalsize

%%%%%%%%%%%%%%%%%%%%%%%%%%%%%%%%%%%%%%%%%%%

%%%%%%%%%%%% Date %%%%%%%%%%%%%%%%%%%%%%%%%%

\par
\bigskip
\par
\hfill{\sl January 1999}
\par
\medskip
\par\rm

%%%%%%%%%%%%%%%%%%%%%%%%%%%%%%%%%%%%%%%%%%%

%%%%%%%%%%%% Abstract %%%%%%%%%%%%%%%%%%%%%%%%

\noindent
{\bf Abstract:}
We conclude the rigorous analysis of the previous paper \cite{m1}
concerning the relation between the (Euclidean) point-splitting approach
and the local $\zeta$-function procedure to renormalize physical quantities
at one-loop in (Euclidean) QFT in curved spacetime. The case of the stress
tensor is now considered in general $D$-dimensional closed manifolds for
positive scalar operators $-\Delta + V(x)$.
Results obtained formally in previous works (in the case $D=4$ and $V(x) =
\xi R(x) + m^2$) are rigorously proven and generalized.
It is also proven that, in static Euclidean manifolds, the method is
compatible with Lorentzian-time analytic continuations.
It is proven
that, for $D>1$, the result of the $\zeta$ function procedure is the same
obtained from an improved version of the point-splitting method which uses a
particular choice of the term $w_0(x,y)$ in the Hadamard expansion of the
Green function, given in terms of heat-kernel coefficients. This version of
the point-splitting procedure works for any value of the field mass $m$.
If $D$ is even, the result is affected by an arbitrary one-parameter class of
(conserved in absence of external source) symmetric tensors,
dependent on the geometry locally, and it
gives rise to the general correct trace expression containing the
renormalized field fluctuations as well as the conformal anomaly term.
Furthermore, it is proven that, in the case $D=4$ and $V(x) = \xi R(x) + m^2$,
the given procedure reduces to the Euclidean version of Wald's improved
point-splitting procedure  provided the
arbitrary mass scale present in the $\zeta$ function is chosen opportunely.
It is finally argued that the found point-splitting method should work
generally, also dropping the hypothesis of a closed manifold, and not depending
on the $\zeta$-function procedure.
This fact is indeed checked in the Euclidean section of Minkowski spacetime
for $A = -\Delta + m^2$ where the method gives rise to the correct Minkowski
stress tensor for $m^2 \geq 0$ automatically.
\par

 \rm

%%%%%%%%%%%%%%%%%%%%%%%%%%%%%%%%%%%%%%%%%%%%%%%%

%%%%%%%%%%%%% PACS numbers %%%%%%%%%%%%%%%%%%%%%%%%%
%\noindent{\sl PACS number(s):\hspace{0.3cm}
%04.62.+v, 04.70.Dy}
%\par
%\bigskip
%\rm

%%%%%%%%%%%%%%%%%%%%%%%%%%%%%%%%%%%%%%%%%%%%%%%%

%%%%%%%%%%%%%%%%  Article text %%%%%%%%%%%%%%%%%%%%%%%

\section*{I. Introduction.}

In a previous paper \cite{m1}, we have considered the relationship 
between the $\zeta$-function and the point-splitting procedures 
in renormalizing some physical quantities: effective Lagrangian, 
effective action and field fluctuations. The more interesting quantity,
namely, the stress tensor, is the object of the present paper.
The aim of this paper is hence twofold. First we want to give a rigorous 
mathematical foundation as well as a generalization of several propositions 
contained in \cite{moa} where they have been stated without rigorous proof. 
This is a quite untrivial task because it involves an 
extension of the heat kernel theory considering the derivatives of its 
usual ``asymptotic'' expansion. As we shall see shortly, this is the core of 
all the analyticity properties
of the generalized tensorial $\zeta$ functions involved in the stress-tensor
renormalization procedure. Second, we want to study the relation between
our technique and the more usual point-splitting  procedure in deep.
This is another open issue  after the appearance of \cite{moa}. 
We know, through practical examples,
that these two approaches agree essentially in several concrete cases, 
but up to now, no general proof of this fact has been given. Anyhow,
it was conjectured by  Wald \cite{wald78} 
that, in general, these two approaches should lead to the same results.
The extension of the $\zeta$-function approach to the stress tensor
 has been introduced in \cite{moa} formally,  this paper contains a proof of
mathematical consistence of the method in a generalized case as well as 
a general proof of the agreement between the two approaches  
under our hypotheses on the manifold and the field 
operator.

It is a well-known fact that the point-splitting procedure 
faces some difficulties in the case of a field which is massless; Indeed, 
in such a case, one cannot make use of the Schwinger-DeWitt algorithm 
to fix $w_0$ in the 
Hadamard expansion \cite{wald78,bd} and, at least in the massless 
conformally coupled case, the point-splitting procedure has been  improved
 in order to get both the conformal anomaly and the conservation of
the renormalized stress tensor \cite{wald78}. Recently,  Wald 
has argued that such an improved procedure, which picks out $w_0\equiv 0$,
can be generalized in more general cases \cite{wald94}.
Differently from the point-splitting procedure,
 the local $\zeta$ function approach seems to work 
without to distinguish between different values of mass and coupling
with the curvature. This fact makes more intriguing the issue of the 
relation between the two procedures.

This paper is organized as follows. In the  first part, we shall recall 
the main features of  the classical theory of the stress tensor to the reader
and
we shall introduce the main ideas concerning the renormalization of the stress
tensor via $\zeta$ function.
In a second part, first  we shall  develop further the 
 theory of the heat-kernel expansion in order to build up the 
theory of the $\zeta$ function of the stress tensor.   
All the work is developed in a closed $D$-dimensional 
manifold for a quite general Euclidean motion operator of a real scalar 
field. 
Successively, we shall state
and prove several theorems concerning generalizations of several mathematical
conjecture employed in \cite{moa}. 
The final part of this work is devoted to investigate the relation
between the two considered techniques and to introduce a generalized
point-splitting procedure. Indeed, within that part, we shall give 
a proof of the agreement of the two approaches,
introducing  an improved point-splitting procedure which is quite similar
and generalizes  that  pointed out in \cite{wald78,wald94}.
We shall see that our prescription 
gives all the expected result (it gives a the trivial 
stress tensor in Minkowski spacetime, the conformal anomaly and a conserved
stress tensor in general, producing agreement with the result of the 
field fluctuations renormalization). 
A final summary ends the work. In the final appendix, the proof of 
some theorems and lemmata is reported.

\section*{II. Preliminaries.}

Within this section, we state the general mathematical 
hypotheses we shall deal with and, very quickly, 
we review 
 the main physical ideas 
concerning the classical stress tensor and its one-loop renormalization 
via point-splitting \cite{bd,wald94,fu}
 and via local $\zeta$ function \cite{moa}. 

We  assume all the definitions and theorems given in the previous paper 
\cite{m1} and we shall refer to those definitions and theorems throughout 
all parts of this work.

\subsection*{A. General hypotheses and notations.}
 
The hypotheses we shall deal with in this work are the same of 
the work \cite{m1}.
 Therefore, from now on, ${\cal M}$ is
 a Hausdorff, connected, oriented,  $C^{\infty}$ Riemannian $D$-dimensional
  manifold.     We suppose also that ${\cal M}$ is compact without 
  boundary (namely  is ``closed'').  
Concerning the operators, 
we shall consider real elliptic differential operators 
with the  Schr\"{o}dinger  form "Laplace-Beltrami operator plus 
potential"
\begin{eqnarray}
A' = - \Delta 
 + V \:\: : \:\:
 C^{\infty}({\cal M}) \rightarrow L^2({\cal M}, d\mu_g)  \label{d}
\end{eqnarray}
where, locally, $\Delta = \nabla_a\nabla^{a}$,
and  $\nabla$ means the covariant derivative associated to the metric
connection, $d \mu_{g} $ is the Borel measure induced by 
the metric, and $V$ is a {\em real} 
function  belonging to  $C^{\infty}({\cal M})$.
We assume that $A'$ is bounded below by some $C\geq 0$ (namely
$A$ is positive). (See sufficient
conditions in \cite{m1}). These are the {\bf general hypotheses} which we shall
refer to throughout the paper.

Moreover, in the most part of the theorems,  we shall use also the fact that 
that
the {\em injectivity radius} 
of the manifold $r$ is strictly positive in closed manifolds 
(see \cite{m1} for further comments on this point).

As general  remarks, we remind the reader that, as in the previous work,
 "holomorphic" and  "analytic" are synonyms  throughout this paper, 
natural units $\hbar=c=1$ are used and the symbol $A$ indicates the
 only self-adjoint extension of the  essentially self-adjoint operator $A'$.
 In the practice, as seen in \cite{m1},
  $A$ coincides with the Friedrichs self-adjoint 
 extension of $A'$.  $R$  indicates the scalar curvature. Moreover, the symbol
$\sigma(x,y)$ means one half the squared geodesical distance of $x$
from $y$ which is continuous everywhere and $C^\infty$  in any
 convex normal neighborhood.

Concerning derivative operators, we shall employ the notations in a fixed
local coordinate system,
\begin{eqnarray}
D^{\alpha}_x := \frac{\partial^{|\alpha|} }{\partial x^{1\alpha_1} \cdots 
 \partial x^{D\alpha_D} }\vert_x
\end{eqnarray}
where the {\em multindex} $\alpha$ is defined by
 $\alpha :=(\alpha_1,\cdots,\alpha_D)$,
 $\alpha_i \geq 0$ is any natural number ($i=1,\cdots,D)$ and 
 $|\alpha| := \alpha_1+ \cdots + \alpha_D $. 
Moreover, $n_k$ will indicate the multindex $(0,\cdots 0,n,0,\cdots0)$
where the only nonvanishing number is $n\in \N$ which takes the $k$th
position.

 Concerning the definitions 
 of the metric-connection symbols and curvature tensors, we shall follows the 
 notations and conventions employed in \cite{m1} which are the same 
employed in \cite{fu}, either for Riemannian
 or Lorentzian signature.

\subsection*{B. Physical background and classical definitions.}

All quantities related to $A'$ we have  considered in the previous 
work \cite{m1} and the averaged stress tensor 
we  consider here, for $D=4$,
 appear in (Euclidean) QFT in curved  
background and concern the theory of quasifree scalar fields.
In several concrete cases of QFT, the form of $V(x)$ 
is $ m^2 + \xi R(x)$
where $m^2$ is the square 
mass of the considered field, $R$ is the scalar curvature of 
the  manifold, $\xi$ is a real parameter.
As usual the  {\em conformal coupling}
 is defined by \cite{bd,fu,wald84}
\begin{eqnarray}
\xi_{D}:= (D-2)/[4(D-1)] \label{cconformal}\:.
\end{eqnarray}

Similarly to the physical quantities considered in the previous work, 
also the stress-tensor is formally 
obtained from the Euclidean functional integral
\begin{eqnarray}
Z[A'] := \int {\cal D} \phi \: e^{-\frac{1}{2} 
\int_{\cal M} \phi A'[{\bf g}] \phi \: d\mu_g}   =:  
e^{-S_{\scriptsize\mbox{eff}}[A']}\:.
\label{integral}
\end{eqnarray} 
$S_{\scriptsize \mbox{eff}} $ is the  
(Euclidean) effective action of the field.
(Here, we use the 
opposite
sign conventions in defining the effective action and thus the stress tensor,
with respect the conventions employed in \cite{moa}. 
Our conventions are the same used in \cite{wa}.) \\
The integral above can be considered as a partition 
function of a field
in a particular  quantum state corresponding to a canonical ensemble 
\cite{wa,ha}.
The direct physical interpretation 
as a partition function should work
provided the manifold has a  static Lorentzian section obtained
by analytically continuing some global temporal coordinate $x^{0}=\tau$ of 
some global chart 
into imaginary values $\tau \rightarrow it$
and considering (assuming that they exist) 
the induced  continuations of the metric and relevant quantities.
It is required also that $\partial_\tau$ is a global 
Killing field of the Riemannian manifold generated by  an isometry
group $S_{1}$. Finally it is required that $\partial_{\tau}$ 
 can be  continued into a
(generally local) time-like Killing field $\partial_{t}$ in the Lorentzian
section (see \cite{ha} and \cite{wa}). Then one assumes that 
$k_{B}\beta$ is the inverse of the 
temperature of the canonical ensemble quantum state, $\beta$ being the 
period of the coordinate $\tau$. The limit case of 
vanishing temperature is also considered 
and in that case  the manifold cannot  be compact. 
Similar interpretations hold for the (analytic continuations of)
 the stress tensor.

Formally \cite{bd,wa,ha,ze,el} we have
$Z[A',{\bf g}] := \left[\mbox{det}\left(\frac{A'}{\mu
^{2}} \right)\right]^{-1/2}$
where  our definition of the determinant of the operator
$A'$ is given by the $\zeta$ function approach \cite{bd,wa,ha,ze,el}
 as pointed out in the 
previous paper \cite{m1}. The scale $\mu^{2}$ 
present in the determinant is necessary for dimensional reasons    
\cite{ha} and plays a central r\^{o}le in the $\zeta-$function 
interpretation of the determinant  and in the consequent theory.
Such a scale introduces an ambiguity which remains in the finite 
renormalization parts of the renormalized quantities and, dealing with
the renormalization of the stress tensor within the
semiclassical approach to the quantum gravity,
it determines the presence of
quadratic-curvature terms in effective Einstein's equations \cite{moa}.
Similar results  are discussed in \cite{wald78,bd,wald94,fu}
employing other renormalization procedures (point-splitting). 

Coming to the (Euclidean) {\em classical} stress-tensor $T_{ab}(x)$, 
it is defined (e.g. see \cite{wald84}) as the locally quadratic form of the 
field obtained by the usual functional derivative once the field 
$\phi$ is fixed
\begin{eqnarray}
T_{ab}[\phi,{\bf g}](x) :=  
\frac{2}{\sqrt g} \frac{\delta} 
{\delta g_{ab}(x)} \left( \frac{1}{2} \: \int_{\cal I} \sqrt{g(x)}
 \phi A'[{\bf g}]
 \phi \: d^Dx \label{ca}
\right)
\:. \label{genstress}
\end{eqnarray}
This functional derivative can be 
 rigorously understood in terms of a  
G\^{a}teaux  derivative for functionals on
real $C^{\infty}({\cal I})$ {\em symmetric} tensor fields $g_{ab}$ and 
the integration above is performed in the open set ${\cal I}$ containing $x$
where the considered coordinate system is defined. 
(\ref{genstress}) means that, and this is the rigorous 
definition
 of the {\em symmetric} 
tensor field $T_{ab}[\phi,{\bf g}](x)$, 
for any $C^{\infty}$ {\em symmetric} tensor field 
 $h_{ab}$ with compact support contained in ${\cal I}$
\begin{eqnarray}
2 \frac{d\:\:\:}{d\alpha}|_{\alpha=0}
S_{\cal I}[{\bf g}+\alpha {\bf h}]   = 
\frac{1}{2} \int_{\cal I} \sqrt{g(x)}
 T_{ab}[\phi,{\bf g}](x) h^{ab}(x) d^Dx
\label{stress222}\:,
\end{eqnarray} 
where
\begin{eqnarray}
S_{\cal I}[\phi,{\bf g}] := \frac{1}{2} \: \int_{\cal I} 
\sqrt{g(x)}\phi A'[{\bf g}] \phi 
\: d^Dx  \:.
\end{eqnarray}
In the case  
\begin{eqnarray}
A'=  - \Delta + m^2 + \xi R(x) + V'(x)\:,
\end{eqnarray}
$V'$ being a $C^{\infty}$ function which does not depend on the metric,
 a direct computation of $T_{ab}(x)$
through this procedure gives 
\begin{eqnarray}
T_{ab}[\phi,{\bf g}](x) &=& \nabla_{a}\phi(x) \nabla_b \phi(x) - \frac{1}{2} 
g_{ab}(x)\left[\nabla_{c}\phi(x) \nabla^c \phi(x) + \left( m^{2}
 + V'(x)\right) \phi^{2}(x)\right] \nonumber \\
&+& \xi \left[ \left(  R_{ab}(x) - \frac{1}{2}g_{ab}(x) 
R(x)\right)\phi^{2}(x)  + g_{ab}(x) \nabla_{c} \nabla^c  \phi^{2} -
\nabla_{a}\nabla_{b} \phi^{2}(x)\right] \label{stresspratico} 
\end{eqnarray}
As well known, $T_{ab}$ given in (\ref{stresspratico}) 
and evaluated for a particular $\phi$
 is {\em conserved} ($\nabla_{a}T^{ab}\equiv  0$) provided $\phi$
is a sufficiently smooth (customary $C^{\infty}$) 
solution of the Euclidean motion, namely,
 $A'\phi \equiv  0$, 
and $V' \equiv 0$. More generally for solution of Euclidean
motion, in local coordinates and for any point $x\in {\cal M}$ one finds
\begin{eqnarray}
\nabla_{a}T^{ab}[\phi,{\bf g}](x) = - \frac{1}{2} \phi^2(x) \nabla^b V'(x)
\label{conservation}\:.
\end{eqnarray}
Another important classical property is the following one.
Whenever the field $\phi$ is massless and conformally coupled 
(i.e. $V'(x) \equiv m^{2} = 0$ and $\xi =\xi_{D})$, the Euclidean action 
$S_{\cal M}$ is invariant under local conformal transformations 
and it holds also 
\begin{eqnarray}
g_{ab}T^{ab}[\phi,{\bf g}](x) = 0 \label{trace}
\end{eqnarray}
everywhere, for smooth fields $\phi$ which are solution of the (Euclidean)
motion equations.
(\ref{conservation}) and (\ref{trace})   
can be checked  for the tensor in (\ref{stresspratico}) directly,
holding our general hypotheses.

 Actually, the requirement of $A'$ 
positive is completely unnecessary for all the definitions 
and results given above 
which hold true in any $C^{\infty}$ Riemannian as well as 
{\em Lorentzian} manifold. In our approach, the Lorentzian stress tensor 
is obtained by analytic
continuation of the Euclidean time as pointed out above.

Passing to the quantum averaged quantities,  following Schwinger
 \cite{sc},  
 the averaged {\em one-loop  stress tensor} for  the quantum state 
determined by the Feynman propagator obtained by the Green function of 
$A'$,
 can be {\em formally} defined by
 \cite{bd,wald94,wa,ha}
\begin{eqnarray}
\langle T_{ab}(x | A') \rangle :=
  \frac{2}{\sqrt{g}} \frac{\delta}{\delta g_{ab}(x)}
S_{\scriptsize\mbox{eff}} =
Z[A',{\bf g}]^{-1}\int {\cal D} \phi \: 
e^{
-\frac{1}{2} \: \int_{\cal M} \phi A'[{\bf g}] \phi \: d\mu_g} 
\: T_{ab}[\phi,{\bf g}](x) \:.
\label{stress}
\end{eqnarray}
It is well-known  that the right hand sides  of
 (\ref{stress}) 
and the corresponding quantity in the Lorentzian section 
are affected by divergences whenever one tries to
compute them by trivial procedures \cite{bd,fu,wa}.
For instance, proceeding as usual (e.g. see \cite{wald94}),
 interpreting the functional integral of $\phi(x) \phi(y)$
 as  a Green function of $A'$ (the 
analytic continuation of the Feynman propagator) $G(x,y)$
and then defining an off-diagonal quantum averaged stress tensor 
\begin{eqnarray}
\langle T_{ab}(x,y) \rangle &=& Z[A',{\bf g}]^{-1}\int {\cal D} \phi  
\:e^{-\frac{1}{2} \: \int_{\cal M} \phi A'[{\bf g}] \phi
d\mu_g }
O_{ab}(x,y) \phi(x)\phi(y) 
\nonumber\\
    &=& O_{ab}(x,y) G(x,y)\:,
\end{eqnarray} 
where $O_{ab}(x,y)$ is an opportune bi-vectorial differential operator
(see \cite{wald94}),
the limit of coincidence of arguments $x$ and $y$,
necessary to get $\langle T_{ab}(x) \rangle$, trivially diverges.
One is therefore forced to remove 
 these divergences {\em by hand} and this is
nothing but the main idea of the {\em point-splitting procedure}. 
Within the point-splitting procedure 
(\ref{conservation})
is requested also for the quantum averaged stress tensor at least in the case 
$V'\equiv 0$. Conversely,
the property (\ref{trace}) generally does not hold  in the 
case of a conformally coupled massless field: a {\em conformal 
anomaly} appears \cite{wald78,bd,wald94,fu}.

Another approach to interpret the left hand side of (\ref{stress}) in 
terms of local $\zeta$ function was introduced in \cite{moa} without 
rigorous mathematical discussion. Anyhow, this  approach has produced
correct results and agreement with point-splitting procedures
 in several concrete cases \cite{moa,dm} and it has 
pointed out a strong self-consistence
and a general agreement \cite{moa}
with the general axiomatic theory of the stress tensor renormalization built up
by Wald  \cite{wald94}. (It is anyway worth stressing that 
Wald's axiomatic approach concerns the Lorentzian theory and thus
any comparison involves an analytical continuation of the Euclidean 
theory. In such a way all the issues related to the locality of the 
theory cannot be compared directly with the general $\zeta$-function
 approach.)
 Moreover, differently from the known point-splitting techniques,
 no difficulty arises
dealing with the case of a massless conformally coupled field.
  
Similarly to the cases treated in 
the previous work \cite{m1},  the  {\em definition}
of the formal quantity 
in the left hand side of (\ref{stress})  
\cite{moa}  given in terms of
$\zeta$ function  and heat kernel 
contains an implicit {\em infinite renormalization} procedure 
in the sense that the result is finally free from divergences.

\subsection*{C. The key idea of the $\zeta$-function regularization of the 
stress tensor.}

The key idea of  $\zeta$-function regularization of the 
stress tensor concerns the extension of the use of the $\zeta$ function from
the effective action to the stress tensor employing some 
manipulations of the series involved in the  $\zeta$-function technique. 
We remind the reader that formally one has \cite{ha,wa,m1}
\begin{eqnarray}
S_{\mbox{eff}}[A]_{\mu^2} =  \frac{1}{2}\frac{d\:}{ds}|_{s=0}
\left\{ - {\sum_{j\in \N}}' \left(\frac{\lambda_j}{\mu^2}\right)^{-s}\right\}
\label{s1}\:.
\end{eqnarray}
$\mu$ is the usual arbitrary mass scale necessary for dimensional reasons.
Actually, the identity above holds true in the sense of the analytic 
continuation. Then, one can try to give some meaning to
the following formal passages 
\begin{eqnarray}
\langle T_{ab}(x) \rangle_{\mu^2}
&=& \frac{2}{\sqrt{g(x)}}\frac{\delta}{\delta g_{ab}(x)}
 \frac{1}{2}\frac{d\:}{ds}|_{s=0}
\left\{ - {\sum_{j\in \N}}' \left(\frac{\lambda_j}{\mu^2}\right)^{-s}\right\}
\nonumber\\
&=& \frac{1}{2}\frac{d\:}{ds}|_{s=0} \left\{  - {\sum_{j\in \N}}' 
\frac{2}{\sqrt{g(x)}}\frac{\delta}{\delta g_{ab}(x)}
\left(\frac{\lambda_j}{\mu^2}\right)^{-s}\right\}\nonumber \\
& = & \frac{1}{2}\frac{d\:}{ds}|_{s=0} \left\{ \frac{s}{\mu^2}  \
{\sum_{j\in \N}}' 
\left(\frac{\lambda_j}{\mu^2}\right)^{-(s+1)}
\frac{2}{\sqrt{g(x)}}\frac{\delta \lambda_j}{\delta g_{ab}(x)} 
\right\}\:.\label{s2}
\end{eqnarray}
The functional derivative of $\lambda_j$ has been computed
in \cite{moa}, at least formally. 
The passages above are mathematically incorrect most likely,
anyhow, in \cite{moa} it was conjectured that the series in the last
line of (\ref{s2}) converges and it can be analytically continued into a
regular function $Z_{ab}(s,x|A/\mu^2)$ in a  neighborhood
of $s=0$. Then, one can {\em define} the renormalized 
averaged one-loop stress tensor
as 
\begin{eqnarray}
\langle T_{ab}(x) \rangle_{\mu^2} :=  \frac{1}{2} \frac{d\:\:}{ds}|_{s=0}
Z_{ab}(s,x|A/\mu^2) \label{s3}\:.
\end{eqnarray}
The explicit form of $Z_{ab}$ found in \cite{moa} following the route above was
\begin{eqnarray}
Z_{ab}(s,x| A/\mu^2) = 2\frac{s}{\mu^2} \zeta_{ab}(s + 1,x|A/\mu^2) + 
s g_{ab}(x) \zeta(s,x| A/\mu^2) \label{Zetas}
\end{eqnarray}
 where $\zeta_{ab}(s,x|A/\mu^2)$ is the analytic continuation of the 
series
\begin{eqnarray}
{\sum_{j\in \N}}' \left( \frac{\lambda_j}{\mu^2}\right)^{-s}
T_{ab}[\phi_j,\phi^*_j, {\bf g}](x) \label{s4}\:,
\end{eqnarray}
and
\begin{eqnarray}
T_{ab}[\phi,\phi^*, {\bf g}](x)
 := -\frac{2}{\sqrt{g(x)}}\frac{\delta\:\:}{\delta g_{ab}(x)}
\frac{1}{2} \int_{\cal I} \phi A'[{\bf g}] \phi^* d\mu_g \label{s5}\:
\end{eqnarray}
However, 
no proof of the convergence of the series above  was given in \cite{moa}
for the general case, but the method was checked in concrete cases finding  
that the series above converges really as supposed.
In \cite{moa}, it was 
showed  also that, assuming reasonable mathematical properties of the 
involved functions, 
 this approach in 
for four-dimensional operators $A' = -\Delta + \xi R(x) + m^2 $ should
 produce  a stress tensor which is conserved and gives rise 
to the conformal anomaly. 
In \cite{moa}, it was also (not rigorously) proven that 
the ambiguity arising from the presence
of the arbitrary scale $\mu^2$ gives rise to conserved geometric
terms added to the stress tensor, in agreement with Wald's axioms.

We expect that the not completely 
rigorous procedures employed
in \cite{moa} make sense provided the usual heat-kernel
``asymptotic'' expansion at $t\rightarrow 0$
 can be derived in the variables which range in the manifold
producing a similar expansion (this result is
not  trivial at all) 
 and  provided the series (\ref{s4}) can be derived 
under the symbol of summation (also this fact is not so obvious). 
Therefore, in the next parts of this work,  we shall investigate 
also similar issues before we prove and generalize all the results found 
in \cite{moa}.

\section*{III.  The local $\zeta$-function and the one-loop stress tensor.}

In this part and within our general hypotheses,
 we develop a rigorous theory of the 
$\zeta$ function of the stress tensor and give a rigorous proof of some
 properties of particular tensorial $\zeta$ functions introduced in \cite{moa}.

The first subsection is devoted to generalize some properties of
the heat-kernel concerning the smoothness of several heat-kernel
expansions necessary in the second subsection.

\subsection*{A. The smoothness of the heat-kernel expansion and the
 $\zeta$ function.}

A first very useful result, which we state in the form of a lemma,
concerns the smoothness of the heat-kernel
expansion for $t\rightarrow 0$ 
({\bf Theorem 1.3} of \cite{m1}) and the possibility of 
deriving term by term such an ``asymptotic expansion''.

Before to state the result it is worth stressing that, in the trivial case
$|\alpha|=|\beta|= 0$, 
the statement of the lemma below and the corresponding proof
include the point $(a2)$ in {\bf Theorem 1.3} of  \cite{m1} 
given without proof there.\\

\noindent {\bf Lemma 2.1.} {\em Let us assume our general
 hypotheses on ${\cal M}$
and $A'$.} {\em For any  
$u \in {\cal M}$ there is an open neighborhood   $I_{u}$  
centered on $u$
such that,  for any
local 
coordinate system defined therein,
for any couple of points $x,y \in I_u$, 
for any couple of multindices $\alpha,\beta$
 and for any integer $N>D/2 + 2|\alpha| +2|\beta|$ ($D/2+2$ if 
$|\alpha| = |\beta| = 0 $) the heat-kernel expansion} (a) {\em of} 
{\bf Theorem 1.3} {\em in} \cite{m1}
{\em can be derived term by term obtaining 
($\eta\in (0,1)$ is fixed arbitrarily
as usual),}
\begin{eqnarray}
D^\alpha_x D^\beta_y K(t,x,y) &=&
D^\alpha_x D^\beta_y \left\{ \frac{e^{-\sigma(x,y)/2t}}{(4\pi t)^{D/2}}
\sum_{j=0}^{N}
a_j(x,y|A)t^j \right\} \nonumber\\
&+& \frac{e^{-\eta\sigma(x,y)/2t}}{(4\pi t)^{D/2}}
 t^{N-|\alpha|-|\beta|} O_{\eta,N}^{(\alpha,\beta)}(t;x,y) \label{first}
\end{eqnarray}
{\em where the derivatives are computed in the common coordinate system
given above and the function $(t,x,y) \mapsto 
O_{\eta,N}^{(\alpha \beta)}(t;x,y) $ 
belongs to $C^0([0, +\infty)\times I_u \times I_u)$ at least, 
and for any positive constant $K_{\eta,N}^{(\alpha,\beta)}$ and
  $0\leq t <K_{\eta,N}^{(\alpha, \beta)} $,  one gets
\begin{eqnarray}
| O_{\eta,N}^{(\alpha, \beta)}(t;x,y)| < M_{K_{\eta,N}^{(\alpha, \beta)}}
 |t| 
\end{eqnarray}
 $M_{K_{\eta,N}^{(\alpha,\beta)}}$ being a corresponding 
 positive constant not dependent on $x,y 
\in I_u$ and $t$.}\\

\noindent {\em Proof.}
See {\bf Appendix} $\Box$.\\

The next lemma concerns the possibility of interchanging the operators
$D_x^\alpha, D_y^\beta$ with the symbol of series in the eigenvector
expansion for the heat kernel given in 
(b) of {\bf Theorem 1.1}  in \cite{m1}.\\

\noindent {\bf Lemma 2.2.} {\em Within our hypotheses on {\cal M} and $A'$,
the eigenvector expansion of the heat kernel given in} (b) {\em of}
 {\bf Theorem 1.1} {\em in} \cite{m1}
\begin{eqnarray}
K(t,x,y|A) = \sum_{j=0}^{\infty} e^{-\lambda_jt} \phi_j(x) \phi^*_j(y)
\end{eqnarray}
{\em 
where $t\in (0, +\infty)$, $x,y\in {\cal M}$ and the real numbers $\lambda_j$ 
($0\leq \lambda_0\leq \lambda_1, \leq \lambda_2, \leq 
\cdots $) are the eigenvalues
of $A$ with corresponding orthogonal 
normalized eigenvector $\phi_j$, can be 
derived in
$x$ and $y$ passing the derivative operators under the symbol of series.
Indeed, in a coordinate system defined in a
 sufficiently small 
neighborhood $I_u$ of any point $u\in {\cal M}$, for $x,y\in I_u$,
for $t\in (0,+\infty)$ and for any couple of multindices $\alpha,\beta$
\begin{eqnarray}
D_x^{\alpha} D_y^{\beta}
 K(t,x,y|A) 
= \sum_{j=0}^{\infty} e^{-\lambda_j t} D_x^{\alpha}\phi_j(x) 
 D_y^{\beta}\phi^*_j(y)\:. \label{pass}
\end{eqnarray}
Moreover, for any $T>0$  the following upper bounds  hold
\begin{eqnarray}
| e^{-\lambda_j t }D_x^\alpha \phi_j(x) D_y^\beta \phi^*_j(y)|
&\leq& P_T^{(\alpha,\beta)}
 e^{-\lambda_j (t-2T)}\:,\\
|D_x^{\alpha} D_y^{\beta}
 K(t,x,y|A) - D_x^{\alpha} D_y^{\beta}
 P_0(x,y|A)| &\leq& P_T^{(\alpha,\beta)} 
{\sum_{j\in \N}}' 
e^{-\lambda_j (t-2T)}\label{ricorda}\\
&\leq& Q_T^{(\alpha,\beta)}\: e^{-\lambda (t -2T)} \label{ricorda'}
\end{eqnarray}
where $x,y\in I_u$ and $t\in (2T, +\infty)$, $P_T^{(\alpha,\beta)}$ and
$Q_T^{(\alpha,\beta)}$ are  positive
constants which do not depend on $t,x,y$, $P_0(x,y|A)$ is the integral kernel
of the projector onto the kernel of $A$, $\lambda$ is the value
of the first strictly positive eigenvalue, the prime on the 
summation symbol indicates that the summation on the vanishing eigenvalues
is not considered, and finally
\begin{eqnarray}
P_T^{(\alpha,\beta)} =
 \left[\sup_{x\in \bar{I}_u}
 ||D_x^\alpha K(T,x,\:.\:|A)||_{L^2({\cal M},d\mu_g)}
\right]
 \left[\sup_{y\in \bar{I}_u}
 ||D_y^\beta K(T,\:.\:, y |A)||_{L^2({\cal M},d\mu_g)}
\right] \label{PQ}
\:;
\end{eqnarray}
Therefore, the convergence of
the series in} (\ref{pass}) {\em is absolute in uniform sense for
$(t,x,y)$ belonging  in any set
$[\gamma,+\infty)\times I_u \times I_u$, $\gamma>0$.}\\

\noindent {\em Proof.}  See {\bf Appendix}. $\Box$\\

\noindent {\em Remark.} The right hand side of (\ref{ricorda})
can be also written down as 
\begin{eqnarray} 
 P_T^{(\alpha,\beta)}\: 
\int_{\cal M}d\mu_g(z) \left\{ K(t-2T,z,z|A)
-P_0(z,z|A)\right\}  = P_T^{(\alpha,\beta)}\:
 Tr \left\{ K_{(t-2T)} - P_0\right\} \label{ricorda''}\:.
\end{eqnarray}

The  two lemmata above enable us to state and prove a theorem concerning
the derivability of the $\zeta$ function.
First of all let us give some definitions (in the following we shall refer to 
 {\bf Definition 2.1} and {\bf Definition 2.2}
 in \cite{m1}).\\

\noindent {\bf Definition 2.1.} 
{\em Let us assume our general hypotheses on ${\cal M}$ and $A'$.
 Fixing a sufficiently small neighborhood $I_u$ 
of any point $u\in {\cal M}$, considering a coordinate system 
defined in $I_u$ and choosing a couple of multindices $\alpha,\beta$,
 the {\bf off-diagonal 
derived local $\zeta$ function} of the operator $A$
is defined for $x,y\in I_u$, $Re$ $s>D/2 + |\alpha| +|\beta|$ 
as 
\begin{eqnarray}
\zeta^{(\alpha,\beta)}(s,x,y|A/\mu^2) := 
D_x^\alpha D_y^\beta \zeta(s,x,y|A/\mu^2) \label{semplice} 
\end{eqnarray}
provided the right hand side exists, where both derivatives are computed 
in the coordinate system defined above.}\\

\noindent {\bf Definition 2.2.} 
{\em Let us assume our general hypotheses on ${\cal M}$ and $A'$.
 Fixing a sufficiently small neighborhood $I_u$ 
of any point $u\in {\cal M}$, considering a coordinate system 
defined in $I_u$ and choosing a couple of multindices $\alpha,\beta$,
 the {\bf derived local $\zeta$ function} of the operator $A$
is defined for $x \in I_u$, $Re$ $s>D/2 + |\alpha| +|\beta|$ 
as 
\begin{eqnarray}
\zeta^{(\alpha,\beta)}(s,x|A/\mu^2) := 
\left\{ 
D_x^\alpha D_y^\beta \zeta(s,x,y|A/\mu^2)\right\}_{x=y}
\label{semplice'} 
\end{eqnarray}
provided the right hand side exists, where both derivatives are computed 
in the coordinate system defined above.}\\

\noindent {\em Remark.} The use of a common coordinate system either
for $x$ and $y$ is essential in these definitions. \\

The following theorem proves that the given definitions make sense.\\

\noindent {\bf Theorem 2.1.}
{\em Let us assume our general hypotheses on ${\cal M}$ and $A'$.
 The local off-diagonal $\zeta$ function of the operator $A$
defined for $x,y\in {\cal M}$, $Re$ $s>D/2$,}
 {\em  $\mu>0$ ($\mu$ being a constant with the dimension of a mass)
\begin{eqnarray}
\zeta(s,x,y|A/\mu^2) = \frac{1}{\Gamma(s)} \int_0^{+\infty}d(\mu^2t)\:
(\mu^2t)^{s-1} \left\{ K(t,x,y|A) -P_0(x,y|A)
\right\}\: \label{zeta}
\end{eqnarray}
can be derived in $x$ and $y$ under the symbol 
of integration in a common 
coordinate system defined in a sufficiently small neighborhood $I_u$ of any
point $u\in {\cal M}$, provided $Re$ $s$ is sufficiently large.
In particular,  for any choice of multindices $\alpha,\beta$
and $x,y\in I_u$ it holds}

 (a) {\em for $Re$ $s> D/2 +|\alpha| +|\beta|$
the derived local $\zeta$ functions are well-defined holding
\begin{eqnarray}
D_x^\alpha D_y^\beta
 \zeta(s,x,y|A/\mu^2) =
\frac{1}{\Gamma(s)} \int_0^{+\infty}d(\mu^2t)\:
(\mu^2t)^{s-1} D_x^\alpha D_y^\beta \left\{
 K(t,x,y|A) -P_0(x,y|A) \right\}\label{semplice''}\:.
\end{eqnarray}
Moreover, the right hand side of} (\ref{semplice''}) 
{\em defines a $s$-analytic
function which belongs to $C^{0}(\{s\in \C |
\mbox{ } Re \mbox{ } s> D/2 +|\alpha| +|\beta|\}\times
I_u \times I_u)$ together with all its $s$-derivatives.}

  (b) {\em Whenever $x \neq y$ are fixed  in $I_u$,}\\
(1) {\em the right hand side of} (\ref{semplice''}) {\em can be
analytically continued in the variable $s$ in the whole complex plane.}\\
(2) {\em Varying $s\in\C$ and  $(x,y) \in (I_u\times I_u)-{{\cal D}_{I_u}} $,
the $s$-continued function in} (\ref{zeta})
{\em defines  an everywhere $s$-analytic  function which belongs to
$C^{\infty}(\C\times \{(I_u \times I_u) - {\cal D}_{I_u}\})$ (where
$ {\cal D}_{I_u} := \{(x,y)\in I_u \times I_u | x = y   \}$) and
it holds, $\C\times \{(I_u \times I_u) - {\cal D}_{I_u}\}$, 
\begin{eqnarray}
D_x^\alpha D_y^\beta \zeta(s,x,y| A/\mu^2) = \zeta^{(\alpha,\beta)}
(s,x,y| A/\mu^2) \label{identita'}
\end{eqnarray}
 where the function $\zeta$  in the left hand side and the
function $\zeta^{(\alpha,\beta)}$ in the right hand side are the respective 
$s$-analytic continuations of the initially defined $\zeta$ 
function} (\ref{zeta})
{\em and the right hand side of} (\ref{semplice''}){\em .}\\
(3) {\em Eq.} (\ref{semplice''}) {\em 
holds also when the left hand side is replaced by  the
$s$-continued function $\zeta^{(\alpha,\beta)}$}
 {\em for $Re$ $s>0$, or everywhere provided
 $ D_x^\alpha D_y^\beta P_0(x,y|A)= 0$ in the considered point
$(x,y)$.} 

 (c) {\em Whenever $x=y$ is fixed in $I_u$,}\\
(1) {\em the right hand side of } (\ref{semplice'})  
{\em can be analytically continued in the variable $s$ in the complex plane 
obtaining a meromorphic function with possible poles, which are 
simple poles only, situated in the points}
\begin{eqnarray}
s^{(\alpha,\beta)}_j &=& D/2+|\alpha|+|\beta|-j, \:\:\:\:\:\:\:\:\:\:
 j = 0,1,2,\cdots
\:\:\:\:\:\:\:\:\: \mbox{{\em if}} \:\:\:
D\:\:\: \mbox{{\em is odd;}} \nonumber\\
s^{(\alpha,\beta)}_j &=& D/2+|\alpha|+|\beta|-j,  \:\:\:\: j = 0,1,2,\cdots
D/2 -1 +|\alpha| +|\beta| \:\:\:\: \mbox{{\em if}} \:\:\: D \:\:\:
\mbox{{\em even.}} \nonumber
\end{eqnarray}
{\em These poles and the corresponding residues are the same of the set 
of analytic functions, labeled by the integer $N> D/2+2|\alpha |+2|\beta |$, 
($N>D/2 +2$ if $|\alpha| =|\beta| =0$)}
\begin{eqnarray}
R_N(s,x)_{\mu_0^{-2}} &:=& \frac{\mu^{2s}}{(4\pi)^{D/2}\Gamma(s)}
\sum_{j=0}^{N} \int_0^{\mu^{-2}_0} dt \times \nonumber \\
&\times& \left\{ D_x^\alpha D_y^\beta 
  e^{-\sigma(x,y)/2t} a_j(x,y|A)\right\}_{x=y} t^{s-1-D/2+j} \:,
 \label{formulazza}
\end{eqnarray}
{\em defined for $x\in I_u$ and  $Re$ $s>D/2 +|\alpha| +|\beta|$
and then continued in the $s$-complex plane. $\mu_0$ is an arbitrary strictly
positive mass scale which does not appear  in the residues.}\\
(2){\em Varying $x\in I_u$, 
the $s$ continued function belongs to
$C^0((\C- {\cal P}^{(\alpha,\beta)})\times {\cal M})$ together 
with all its $s$ derivatives,
 ${\cal P}^{(\alpha,\beta)}$ 
 being the  set 
of the actual poles (each for some $x$) among the points  listed above.
Moreover,  for any coordinate $x^k$ and  $(s,x)\in
\{\C- ({\cal P}^{(\alpha,\beta)}
\cup {\cal P}^{(\alpha + 1_k,\beta)}
 \cup {\cal P}^{(\alpha,\beta +1_k)})\}\times I_u$,}
$\frac{\partial}{\partial x^k} 
\zeta^{(\alpha,\beta)}(s,x|A/\mu^2)$ {\em exists, is continuous in $(s,x)$
with all of its $s$ derivatives, analytic in the variable $s$
and}
\begin{eqnarray}
\frac{\partial}{\partial x^k} 
\zeta^{(\alpha,\beta)}(s,x|A/\mu^2)
 = \zeta^{(\alpha+1_k,\beta)}(s,x|A/\mu^2) 
+ \zeta^{(\alpha,\beta + 1_k)}(s,x|A/\mu^2)  
\label{piu'uno}
\end{eqnarray}
{\em where $\zeta^{(\alpha,\beta)}$
is the analytic continuation of the initially defined function} 
(\ref{semplice'}).

(d) {\em For $x,y \in I_u$, the analytic continuations of the right hand 
sides of } (\ref{semplice}) {\em and} (\ref{semplice'})
{\em are well-defined in a neighborhood of $s=0$ and it holds, and the 
result does not  depends on the values of $\mu_0>0$ and $\mu >0$,} 
\begin{eqnarray}
\left[ D_x^\alpha D^\beta_y
 \zeta (s,x,y|A/\mu^2)\right]_{s=0}
+  D_x^\alpha D^\beta_y P(x,y) = \delta_D \delta_{x,y} 
\lim_{s\rightarrow 0} R_N(s,x)_{\mu^{-2}_0}\:,
 \label{finet}
\end{eqnarray}
{\em where $|_{s=0}$ means the analytic continuation from $Re$ $s>
D/2 +|\alpha |+|\beta |$ to $s=0$ of the considered function and $N$
is any integer $> D/2 +2|\alpha| +2|\beta|$  ($D/2$ whenever
 $|\alpha|=|\beta|=0$).  Finally, we have defined 
$\delta_D = 0$ if $D$ is odd and 
 $1$ otherwise, $\delta_{x,y}=0$ if $x\neq y$ or $1$ otherwise.}\\

\noindent {\em Sketch of  Proof.} The proof of this theorem is a
straightforward generalization of the proof of {\bf Theorem 2.2} in
\cite{m1}, so we just sketch this proof.
As in the proof of {\bf Theorem 2.2} in \cite{m1}, 
the main idea is to break off the 
integration in (\ref{zeta}) for $Re$ $s>D/2$ as
\begin{eqnarray}
\zeta(s,x,y|A/\mu^2) &=& \frac{\mu^{2s}}{\Gamma(s)}
\int_0^{+\infty} dt\: t^{s-1} 
\left[ K(t,x,y|A) - P_0(x,y|A) \right] \\
&=&  \frac{\mu^{2s}}{\Gamma(s)} \int_{0}^{\mu_{0}^{-2}} 
\left\{\ldots\right\} + \frac{\mu^{2s}}{\Gamma(s)} 
\int_{\mu_{0}^{-2}}^{+\infty} 
\left\{  \ldots  \right\}\:,  \label{break}
\end{eqnarray}
where  $\mu_{0}>0$ is an arbitrary mass cutoff.
Then one studies the possibility of computing the derivative passing
$D_x^\alpha$ and $D_y^\beta$ under the symbol of integration in both
integrals in the right hand side above.  This is possible provided the absolute
values of the derived integrand are $x,y$ uniformly bounded by integrable 
functions dependent on $\alpha,\beta$ in general, 
for any choice of $\alpha$ and $\beta$. This assures also the continuity 
of the derivatives because the derivatives of the integrands are 
continuous functions. The analyticity in $s$ can be proved by checking
the Cauchy-Riemann conditions passing the derivative under the symbol
of integration once again. The $s$-derivatives of the integrand at any order
can be still proven to be  bounded with the same procedure.
Then the proof deal with similarly  to  the 
proof of {\bf Theorem 2.2} of \cite{m1}. One uses {\bf Lemma 2.2} and
(\ref{ricorda'}) (choosing $2T< \mu_0^{-2}$) in place 
of the corresponding formula (99) of \cite{m1}, 
 to prove that the latter integral in the right hand side of 
(\ref{break}) can be derived under the symbol of integration obtaining 
a $s$-analytic function continuous with all of its $s$ derivatives, 
for $s\in \C$ and $x,y \in I_u$. The former integral can be studied employing
{\bf Lemma 1.1} and, in particular, (\ref{first}).
The requirement $N>D/2+2$ in the expansion in \cite{m1}
 has to be changed $N> D/2 +2|\alpha| +2|\beta|$ in the present case.
 The requirement in the point (a) $Re$ 
 $s> D/2 +|\alpha| +|\beta|$ arises by the term with $j=0$ in the heat
kernel expansion when all the derivatives either in $x$ and in $y$ act
on the exponential  producing a factor $t^{-|\alpha|-|\beta|}$ and posing
$x=y$ in the end.
Eq. (101) and the successive ones of \cite{m1} have to be changed employing
$O^{(\alpha,\beta)}_\eta$ in place of $O_\eta$  and
$t^{s-1+N-D/2 -|\alpha| -|\beta|}$ in place of
$t^{s-1+N-D/2}$. 

The requirement $D_x^\alpha D_y^\beta P_0(x,y|A)=0$ 
in (b3)
is simply due to the divergence of the integral
$ \int_0^{\mu^{-2}_0} dt t^{s-1}$
for $s\leq 0$.\\
(\ref{finet}) is essentially due to the presence of the factor
$1/\Gamma(s)$ in all considered integrals, which vanishes with a simple
zero as $s \rightarrow 0$.
$\Box$\\

\noindent {\em Comments}\\
{\bf (1)} 
The right hand side of  (\ref{finet}), for $x=y$ and when  $D$ is even
has the form
\begin{eqnarray}
\frac{D_x^\alpha D^\beta_y a_{D/2}(x,y|A)  + \cdots }{(4\pi)^{D/2}}
\end{eqnarray}
Where
 the dots indicate a finite number of further terms consisting of derivatives
of product of heat kernel coefficients and powers 
of $\sigma(x,y)$, computed in the
coincidence limit of the arguments. In the case $|\alpha|=|\beta|=0$
this agrees with the found result for the simple local $\zeta$ function
given in \cite{m1}.\\
{\bf (2)}  It is worth noticing that the right hand side of 
(\ref{finet}) proves that the procedures of $s$-continuing 
$D_x^\alpha D_y^\beta \zeta(s,x,y| A/\mu^2)$ and that of 
taking the coincidence
limit of arguments $x,y$ {\em generally do not commute}. 
This means that, understanding both sides in the sense of the analytic
continuation,  in general
\begin{eqnarray}
\zeta^{(\alpha,\beta)}(s,x,y|A/\mu^2)|_{x=y} \neq 
\zeta^{(\alpha,\beta)}(s,x|A/\mu^2) \:.
\end{eqnarray}
Above the coincidence limit is taken {\em after} the analytic continuation.
Obviously, whenever $Re$ $s> D/2 +|\alpha|+ |\beta|$ 
\begin{eqnarray}
\zeta^{(\alpha,\beta)}(s,x,y|A/\mu^2)|_{x=y} =
\zeta^{(\alpha,\beta)}(s,x|A/\mu^2) \:.
\end{eqnarray}
{\bf (3)}  The point (b2) proves that, for $x\neq y$, the Green function 
of any operator $A^n$, $n=0,1,2\cdots$ defined in \cite{m1} via local $\zeta$
function, is $C^{\infty}$ as one could have to  expect. \\

A second and last theorem concerns the possibility to compute the
derived local $\zeta$ functions through a series instead of an integral.\\

\noindent {\bf Theorem 2.2} {\em Within our hypotheses on ${\cal M}$
and $A'$  and $\mu>0$, the (off-diagonal and not)
derived local $\zeta$ function can be computed as the sum of a series.
Indeed, choosing a couple of multindices $\alpha,\beta$, in a common 
coordinate system defined in a sufficiently small 
neighborhood of any  point $u\in
{\cal M}$ one has, in the sense of the punctual convergence
\begin{eqnarray}
\zeta^{(\alpha,\beta)}(s,x,y|A/\mu^2)
&=& {\sum_{j\in \N}}' \left(\frac{\lambda}{\mu^{2}}\right)^{-s}
D_x^\alpha \phi_j(x) D_y^\beta \phi^*_j(y) \label{serie}\\
\zeta^{(\alpha,\beta)}(s,x|A/\mu^2)
&=& {\sum_{j\in \N}}' \left(\frac{\lambda}{\mu^{2}}\right)^{-s}
D_x^\alpha \phi_j(x) D_x^\beta \phi^*_j(x) \label{serie'}\:, 
\end{eqnarray}
provided  $Re$ $s> 3D/2 +|\alpha| +|\beta|$ and $(x,y)\in I_u\times I_u$.}\\

\noindent {\em Proof.}
First of all it is worth stressing that, in the considered domain 
for $s$, the functions are continuous in all variables and 
\begin{eqnarray}
\zeta^{(\alpha,\beta)}(s,x,y|A/\mu^2)|_{x=y} =
\zeta^{(\alpha,\beta)}(s,x|A/\mu^2) \:.
\end{eqnarray}
So we perform our proof in the general case $x\neq y$ and then consider
the coincidence limit of arguments.  Therefore,
from {\bf Theorem 2.1.}, for $Re$ $s> D/2 +|\alpha| +|\beta|$,
one has
\begin{eqnarray}
\zeta^{(\alpha,\beta)}(s,x,y|A/\mu^2) &=& 
\frac{\mu^{2s}}{\Gamma(s)}
\int_0^{\mu_0^{-2}} dt\: t^{s-1} D_x^\alpha D_y^\beta 
\left\{ K(t,x,y|A) -P_0(x,y|A) \right\}\nonumber \\
&+& \frac{\mu^{2s}}{\Gamma(s)}
\int_{\mu_0^{-2}}^{+\infty} dt\: t^{s-1} D_x^\alpha D_y^\beta 
\left\{ K(t,x,y|A) -P_0(x,y|A) \right\} \label{trunc} \:.
\end{eqnarray}
$\mu_0>0$ arbitrarily.
Let us focus attention on the second integral. It can be written also 
\begin{eqnarray}
\frac{\mu^{2s}}{\Gamma(s)}
\int_{\mu_0^{-2}}^{+\infty} dt\: {\sum_{j\in \N}}'t^{s-1} D_x^\alpha 
\phi_j(x) D_y^\beta \phi^*_j(y) e^{-\lambda_j t}\label{uno}\:,
\end{eqnarray}
where we have used {\bf Lemma 2.2}. We want to show that it is possible
to interchange the symbol of series with that of integration. We shall prove 
a similar fact for the other integral in (\ref{trunc}), then  the well-known
formula ($a>0$)
\begin{eqnarray}
a^{-s} = \frac{1}{\Gamma(s)}\int_0^{+\infty} dt\: t^{s-1} e^{-at} 
\label{sempre}
\end{eqnarray}
will complete the proof of the theorem.

To prove the possibility of interchanging the integration with the
summation in the integral (\ref{uno}) it is sufficient to show that
the absolute value of the function after the summation symbol
 is integrable in the measure
$\int dt \times \sum_j $, then Fubini's theorem allows one to interchange
the integrations.
From {\bf Lemma 2.2}, we know that for $t>2T>0$
\begin{eqnarray}
{\sum_{j\in\N}}'
 \: |t^{s-1} D_x^\alpha 
\phi_j(x) D_y^\beta \phi^*_j(y) e^{-\lambda_j t}|
&\leq& P_T^{(\alpha,\beta)} {\sum_{j\in\N}}'
t^{Re\: s-1}e^{-\lambda_j (2t-2T)} \nonumber\\
&\leq& Q_T^{(\alpha,\beta)} t^{Re\:s -1} e^{-\lambda (t-2T)} \:,
\end{eqnarray}
where $\lambda$ is the first strictly positive eigenvalue of $A$.
We choose the constant $T< \mu^{-2}_0/2$. The $t$-integration in $[\mu^{-2}_0,
+\infty)$ of the last line above is finite for any $s\in \C$. Thus, 
a part of
Fubini's theorem prove that the function after the summation symbol
 in (\ref{uno}) is integrable in the
product measure.\\
Let us perform a similar proof for the first integral in the right hand side 
of (\ref{trunc}). It can be written down
\begin{eqnarray}
\frac{\mu^{2s}}{\Gamma(s)}
\int^{\mu_0^{-2}}_0 dt\: {\sum_{j\in \N}}'t^{s-1} D_x^\alpha 
\phi_j(x) D_y^\beta \phi^*_j(y) e^{-\lambda_j t}\label{due}\:.
\end{eqnarray}
We want to show that it is possible to interchange the symbol of series
with that of integration.
 Posing $T=t/4$
we have, for $t\in (0, \mu^{-2}_0]$
\begin{eqnarray}
{\sum_{j\in \N}}'|t^{s-1} D_x^\alpha 
\phi_j(x) D_y^\beta \phi^*_j(y) e^{-\lambda_j t}|
&\leq& P_{t/4}^{(\alpha,\beta)} t^{Re\: s-1} Tr \left\{ K_{t/2} 
-P_0\right\}\label{qfine}\:.
\end{eqnarray}
where  for (\ref{PQ}) 
\begin{eqnarray}
P_T^{(\alpha,\beta)} :=
 \left[\sup_{x\in \bar{I}_u} ||D_x^\alpha K(T,x,\:.\:|A)||\right]
 \left[\sup_{y\in \bar{I}_u} ||D_y^\beta K(T,\:.\:, y |A)||\right] 
\nonumber\:.
\end{eqnarray}
Employing (\ref{first}) of {\bf Lemma 2.1} and taking
 account of the finite volume of the manifold 
one finds  that
there is a positive constant $A$ such that,
for $t\in (0, \mu^{-2}_0] $
\begin{eqnarray}
P_T^{(\alpha,\beta)} \leq A  T^{-D -|\alpha| -|\beta|}\:.
\end{eqnarray}
This is due to the leading order for $t\rightarrow 0$ of the heat-kernel
expansion (\ref{first}).
This upper bound, inserted in (\ref{qfine}) with $T= t/4$,
together with the $x$-integral of the heat-kernel expansion (19)
in {\bf Theorem 1.3} of \cite{m1},  entails
\begin{eqnarray}
{\sum_{j\in \N}}' |t^{s-1} D_x^\alpha 
\phi_j(x) D_y^\beta \phi^*_j(y) e^{-\lambda_j t}|
\leq B
t^{Re\: s-1 - 3D/2 -|\alpha| -|\beta| } \label{qfine'} \:.
\end{eqnarray}
where $B$ is a positive constant.

Dealing with as in previously considered case,
for  $Re$ $s> 3D/2 +|\alpha| +|\beta|$, we can interchange the symbol 
of integral with that of series also in the second integral of (\ref{trunc}),
then (\ref{sempre}) entails the thesis. $\Box$\\

 Notice that, in the case $|\alpha|= |\beta| =0$, the 
convergence of the series (\ref{serie'}) 
arises for $Re$ $s>D/2$ and it is uniform as well-known \cite{m1}.
Actually, our theorem uses a quite rough hypothesis. 
Nevertheless, this is enough for the use we shall make of the theorem above.

Following the way traced out in {\bf 1.3}, we can give a precise definition
concerning the $\zeta$ function of the stress tensor. We shall assume, more
generally than in \cite{moa},
$A' := -\Delta + V$ where $V(x):= m^2 +\xi R + V'(x)$ and $V'$ is real and 
$\in 
C^{\infty}({\cal M})$ does not depend on the metric. 
Moreover, in this paper we consider a general $D$ dimensional manifold
rather than the more physical case $D=4$ studied in \cite{moa}.
Also, as required by our general hypotheses, $A'$
must be positive. It is worth stressing that this does not entails
necessarily $m^2 + \xi R(x) > 0 $ everywhere also when $V'\equiv 0$, and
neither  $V(x)>0$ everywhere in the general case (see \cite{m1}).

\subsection*{B. The $\zeta$-regularized stress tensor and its properties.}

For future convenience, let us define
the  symmetric tensorial field  in a local coordinate system,
\begin{eqnarray}
T_{ab}[\phi,\phi^*, {\bf g}](x)
 := \frac{2}{\sqrt{g(x)}}\frac{\delta\:\:}{\delta g_{ab}(x)}
\frac{1}{2} \int_{\cal I} \phi A'[{\bf g}] \phi^* d\mu_g \label{tab}\:,
\end{eqnarray}
where $\phi \in C^{\infty}({\cal M})$ and the functional derivative has been
defined in {\bf 1.2}. The precise form of 
$T_{ab}[\phi,\phi^*, {\bf g}](x)$ reads in our case
\begin{eqnarray}
T_{ab}[\phi,\phi^*, {\bf g}](x) &=& 
\frac{1}{2}(\nabla_{a}\phi(x) \nabla_b \phi^*(x) +\nabla_{a}\phi^*(x) 
\nabla_b \phi(x) )\nonumber\\
& &- \frac{1}{2} 
g_{ab}(x)\left[\nabla_{c}\phi(x) \nabla^c \phi^*(x) + \left( m^{2}
 + V'(x)\right) |\phi|^{2}(x)\right] \nonumber \\
& & +\xi \left[ \left(  R_{ab}(x) - \frac{1}{2}g_{ab}(x) 
R(x)\right)|\phi|^{2}(x)  + g_{ab}(x) \nabla_{c} \nabla^c  |\phi|^{2}(x)
\right. \nonumber \\ 
& & - \left. \nabla_{a}\nabla_{b} |\phi|^{2}(x)\right] \label{stresspratico'}\:. 
\end{eqnarray}
A few trivial manipulations which make use of $A'\phi_j = \lambda_j \phi_j$
lead us to a simpler form for $T_{ab}[\phi_j,\phi_j^*, {\bf g}](x)$, namely
\begin{eqnarray}
T_{ab}[\phi_j,\phi_j^*, {\bf g}](x) &=& 
\frac{1}{2}(\nabla_{a}\phi_j(x) \nabla_b \phi_j^*(x) +\nabla_{a}\phi_j^*(x) 
\nabla_b \phi_j(x) )\nonumber\\
& &- \xi \nabla_{a} \nabla_{b}  |\phi_j|^{2} + \left(\xi -\frac{1}{4} \right)
g_{ab}(x) \Delta |\phi_j|^2(x) \nonumber\\
& &+ \xi R_{ab}(x)|\phi_j|^2 - \frac{g_{ab}(x)}{2}
\lambda_j |\phi_j|^2(x) \label{tab'}
\:.
\end{eqnarray}

Following the insights given in {\bf 1.3} as well as \cite{moa}, we can give
the following definition.\\

\noindent {\bf Definition 2.3.}
{\em Within our hypotheses on ${\cal M}$ and 
$A' := -\Delta + m^2 +\xi R + V'(x)$ defined above ($m,\xi \in \R$), the 
{\bf local $\zeta$ function of the stress tensor} is the symmetric
tensorial field  defined in local coordinates as} 
\begin{eqnarray}
Z_{ab}(s,x| A/\mu^2) := 2\frac{s}{\mu^2} \zeta_{ab}(s + 1,x|A/\mu^2) + 
s g_{ab}(x) \zeta(s,x| A/\mu^2) \label{Zeta}
\end{eqnarray}
{\em where $\zeta_{ab}(s,x|A/\mu^2)$ is defined  as the sum of 
the series below, in a sufficiently small 
neighborhood $I_u$
of any point $u\in {\cal M}$ and for $Re$ $s> 3D/2 + 2$,
\begin{eqnarray}
{\sum_{j\in \N}}' \left(\frac{\lambda_j}{\mu^2}\right)^{-s} 
T_{ab}[\phi_j,\phi_j^*, {\bf g}](x)
 \label{zetaab}\:,
\end{eqnarray}
and $T_{ab}[\phi_j,\phi_j^*, {\bf g}](x)$ is defined in} (\ref{tab}) {\em and}
(\ref{tab'}) {\em with respect to a base of smooth orthogonal normalized
eigenvector of $A$.}\\

\noindent {\em Comments}\\
{\bf (1)}  The definition given above makes sense since
the relevant series converges for $Re$ $s> 3D/2 + 2$ because of 
 {\bf Theorem 2.2}. Notice that the given definition does not depend
on the base of smooth orthogonal normalized eigevectors of $A$ (take account
that each eigenspace has finite dimension as follows from {\bf Theorem 1.1}
in \cite{m1}) . \\
{\bf (2)}  The fact that the coefficients $Z_{ab}(s,x|A/\mu^2)$ do
define a tensor is a direct consequence of (\ref{tab'}) and (\ref{zetaab}).
This can be trivially proven for $Re$ $s>3D/2 +2$ where one can make use
of the series (\ref{zetaab}) which trivially define a tensor since all
terms of the sum are separately components of a tensor. 
 Then, the proven
property remains unchanged after the analytic continuation for values
of $s$ where the series does not converge.\\
{\bf (3)}  It is worthwhile noticing that the final expression of $T_{ab}$
and thus $Z_{ab}$ self, does not contain either $m^2$ or $V'$ explicitly.\\ 
{\bf (4)}  A definition trivially equivalent to (\ref{Zeta}) (up to analytic 
continuations in the variable $s$) is given by posing directly, for
$Re$ $s> 3D/2 +2$,
\begin{eqnarray}
 Z_{ab}(s,x|A/\mu^2) &=&
s  {\sum_{j\in \N}}' \left\{ \frac{2}{\mu^2}
\left(\frac{\lambda_j}{\mu^2}\right)^{-(s+1)} 
T_{ab}[\phi_j,\phi_j^*, {\bf g}](x) \right. \nonumber\\
& +& \left.  g_{ab}(x)
\left(\frac{\lambda_j}{\mu^2}\right)^{-s} \phi_j(x) \phi_j^*(x)
\right\} 
 \label{Zeta2'}\:.
\end{eqnarray}\\
{\bf (5)}   Due to the uniqueness theorem for analytic 
functions and {\bf Theorem 2.2}, each component of $Z_{ab}$ can be built up,
within opportune regions,
 employing the heat kernel in the fashion of  {\bf Theorem 2.1}.
Following this route, 
by {\bf Theorem 2.1}, one proves trivially that the symetric 
tensorial field  $Z_{ab}(s,x|A/ \mu^2)$ is continuous
together all of its $s$ derivatives as a function of  $(s,x)$. In particular 
each component 
 defines a meromorphic  function of the variable $s$ whenever $x$ is fixed.
Therefore,
 let us consider  the {\em simple } poles which may appear in the $s$-continued
components of $Z_{ab}(s,x|A/\mu^2)$. One has to rewrite each component 
of this tensorial field in terms of simple $\zeta$ functions and
derived $\zeta$ functions via {\bf Theorem 1.1} and {\bf Theorem 1.2}.
 Once this has been done, one sees that the only
functions which really appear in $Z_{ab}$ are $\zeta^{(1_a,1_b)}$,
$\zeta^{(1_b,1_a)}$  
 and $\zeta$, each function  evaluated at $s+1$ (see (\ref{sotto})).
 The simple local $\zeta$ 
function evaluated in $s+1$ admits possible simple poles
in the points $s_j$ with $s_j = D/2 - j -1$ 
where $j= 0,1, \cdots  $ whenever
$D$ is odd, otherwise $j= 0, 1, \cdots D/2 -1$ whenever $D$ is even.
Anyhow, the factor $s$ in (\ref{Zeta}) cancels out the possible pole at $s=0$,
 which may appears in $\zeta(s+1)$ when $D$ is even. 
The functions 
$\zeta^{(1_c,1_d)}(s+1,x|A/\mu^2)$ have been  
considered in {\bf Theorem 1.1} and their possible simple poles
may arise in  $s_j$ with $s_j = D/2 + |1_a| + |1_b| -1 - j
= D/2 -j +1$ and $j= 0,1, \cdots  $ whenever
$D$ is odd, otherwise $j= 0, 1, \cdots D/2 + 1$ whenever $D$ is even.
Once again, because of the factor $s$ in the right hand side of (\ref{Zeta})
any possible simple pole at $s=0$ is canceled out.

The last comment above can be stated into a theorem.\\

\noindent {\bf Theorem 2.3.} {\em In our general hypotheses
on ${\cal M}$ and $A'$}
(a)  {\em each component of 
$Z_{ab}(s,x|A/\mu^2)$ can be analytically
 continued into a meromorphic
function of $s$ whenever $x$ is fixed. In particular,  in the sense of the
analytic continuation, it holds,
for $x$ belonging to 
 a sufficiently small neighborhood of any point $u\in {\cal M}$
\begin{eqnarray}
Z_{ab}(s,x|A/\mu^2) &=& \frac{s}{\mu^2}\left[
\zeta^{(1_a,1_b)}(s+1,x|A/\mu^2) +\zeta^{(1_b,1_a)}(s+1,x|A/\mu^2) 
\right] \nonumber\\
&+& \frac{2s}{\mu^2}\left[\left(\xi - \frac{1}{4}\right)  g_{ab}(x) \Delta 
+ \xi R_{ab}(x) - \xi \nabla_a \nabla_b \right] \zeta(s+1,x|A/\mu^2)
\label{sotto}
\end{eqnarray}}

(b) {\em The possible 
poles of each component of $Z_{ab}(s,x|A/\mu^2)$, which are 
simple poles only, are situated in the points}
\begin{eqnarray}
s_j &=& D/2 -j + 1, \:\:\:\:\:\:\:\:\:\:
 j = 0,1,2,\cdots
\:\:\:\:\:\:\:\:\: \mbox{{\em if}} \:\:\:
D\:\:\: \mbox{{\em is odd;}} \nonumber\\
s_j &=& D/2 -j + 1,  \:\:\:\: j = 0,1,2,\cdots
D/2   \:\:\:\: \mbox{{\em if}} \:\:\: D \:\:\:
\mbox{{\em even.}} \nonumber
\end{eqnarray}

(c)
{\em Varying $x\in I_u$ and $s\in \C$ the $s$-analytically continued symmetric
tensorial field
$(s,x) \mapsto Z_{ab}(s,x|A/\mu^2)$ defines a $s$-analytic tensorial field of 
$C^0((\C-{\cal P})\times I_u)$ together with all its $s$ derivatives, where
${\cal P}$ is the set of the actual poles (each for some $x$ ) among the
points listed above.}\\

\noindent {\em Proof.} Sketched above. $\Box$\\

\noindent 
{\em Remark.} Eq. (\ref{sotto}) could be used as an independent definition
of the $\zeta$ function of the stress tensor. The important point is that
it does not refer to any series of eigenvectors. It could be considered
as the starting point for the generalization of this theory in the case
the spectrum of the operator $A$ is continuous provided the functions
in the right hand side of (\ref{sotto}) are defined in terms of $t$
 integrations of derivatives of the heat kernel.\\

\noindent {\bf Definition 2.4.} {\em In our general hypotheses
on ${\cal M}$ and $A'$  and for 
$x\in I_u$  where $I_u$ is a sufficiently small neighborhood of 
$u\in {\cal M}$, the {\bf one-loop renormalized stress tensor}
is defined in a local coordinate system in $I_u$,
 by the set of functions ($a,b =1,\cdots,D$)
\begin{eqnarray}
\langle T_{ab}(x|A) \rangle_{\mu^2}
 := \frac{1}{2} \frac{d\:}{ds}|_{s=0}
Z_{ab}(s,x|A/ \mu^2) \label{ztensor}\:,
\end{eqnarray}
where the tensorial field $Z_{ab}$ which appears in the right hand side
is is the $s$-analytic continuation of that defined above and
$\mu^2>0$ is any fixed constant with the dimensions of a squared mass.}\\

 We can state and prove the most important properties of 
$\langle T_{ab}(x|A) \rangle_{\mu^2}$
 in the following theorem.
These results generalize  previously obtained results \cite{moa,dm}
for a more general operator $A$ and for any dimension $D>0$.\\

\noindent {\bf Theorem 2.4.}
 {\em In our general hypotheses
on ${\cal M}$ and $A'$, 
the functions $x \mapsto \langle T_{ab}(x|A) \rangle_{\mu^2}$
 defined above satisfy the following properties.}

(a) {\em The functions  $x \mapsto \langle T_{ab}(x|A) \rangle_{\mu^2}$
 ($a,b=1,2, \cdots, D $) define
 a  $C^\infty$ symmetric  tensorial field  on ${\cal M}$.}

(b) {\em This tensor is {\em conserved} for $V'\equiv 0$, and more generally}
\begin{eqnarray}
\nabla^{a} \langle T_{ab}(x|A) \rangle_{\mu^2} = -\frac{1}{2}\langle
\phi^2(x|A)\rangle_{\mu^2}\: \nabla_b V'(x) \label{t1} 
\end{eqnarray}
{\em everywhere in} ${\cal M}$. 

(c){\em For any rescaling $\mu^2 \rightarrow \alpha \mu^2$, where
$\alpha>0$
is a pure  number, one has
\begin{eqnarray}
\langle T_{ab}(x|A) \rangle_{\mu^2} 
\rightarrow \langle T_{ab}(x|A) \rangle_{\alpha\mu^2} =
\langle T_{ab}(x|A) \rangle_{\mu^2} + (\ln \alpha) t_{ab}(x|A) \label{t2}
\end{eqnarray}
where $t_{ab}(x|A)
  = Z_{ab}(0,x|A)/2$, which coincides also with the residue
of the pole
of $\zeta_{ab}(s+1,x|A)$ at $s=0$,
 is a,  conserved for $V'\equiv 0$,  symmetric tensor not dependent
on $\mu$ built up
by a linear combination of product of the metric, 
curvature tensors, $V'(x)$ and their
covariant derivatives evaluated at the point $x$. In general it satisfies
\begin{eqnarray}
\nabla^{a} t_{ab}(x|A) = 
-\delta_D\frac{a_{D/2-1}(x,x|A)}{2(4\pi)^{D/2}}  \nabla_b
 V'(x) \label{aggiunta}\:,
\end{eqnarray}
 where $\delta_D =0$ when
$D$ is odd and $\delta_D = 1$ otherwise.
In terms of
heat-kernel coefficients one has also}
\begin{eqnarray}
 t_{ab}(x|A) &=& \frac{\delta_D}{(4\pi)^{D/2}}
 \left\{a_{D/2-1,(ab)}(x,x|A)  
+ \frac{g_{ab}(x)}{2} a_{D/2}(x,x|A)\right. \nonumber \\
&+& \left. \left[ \left(\xi -\frac{1}{4}\right)g_{ab}(x) 
\Delta
 + \xi R_{ab}(x) -\xi 
\nabla_a \nabla_b  \right] a_{D/2-1}(x,x|A)\right\} \label{twald}\:,
\end{eqnarray}
{\em where we have employed the notations (using the same coordinate system
both for $x$ and $y$)} 
\begin{eqnarray}
a_{j,(ab)}(x,x|A) := \frac{1}{2}\left[\left(\nabla_{(x)a}\nabla_{(y)b} 
+ \nabla_{(y)a}\nabla_{(x)b}\right) a_{j}(x,y|A)\right]|_{x=y}   
\end{eqnarray}

(d) {\em Concerning the trace of $\langle T_{ab}(x|A) \rangle_{\mu^2}$
 one has} 
\begin{eqnarray}
g^{ab}(x)\langle T_{ab}(x|A) \rangle_{\mu^2} &=& \left( 
\frac{\xi_D-\xi}{4\xi_D-1} \Delta - m^2 - V'(x)\right)\langle \phi^2(x|A)
\rangle_{\mu^2} \nonumber\\
&+& \delta_{D}\frac{a_{D/2}(x,x|A)}{(4\pi)^{D/2}} - P_0(x,x|A) \label{t3}\:.
\end{eqnarray}
{\em Above, $\langle \phi^2(x|A)
\rangle_{\mu^2}$ is the value of 
 the averaged quadratic fluctuations of the field computed
by the $\zeta$-function approach} \cite{m1,dm}.\\

(The coefficient $(4\xi_D-1)^{-1}$ above is missprinted in
\cite{dm} where $(2\xi_D)^{-1}$ appears in place of it.)\\

\noindent {\em Sketch of Proof.}
Barring the issue concerning the smoothness, 
the property (a) is a trivial consequence of the corresponding fact for
$Z_{ab}(s,x|A/\mu^2)$ 
discussed in Comment (2) after {\bf Definition 2.3}.
The tensorial field belongs to $C^\infty$ because of the $C^\infty$ smoothness
of the functions $(s,x)\mapsto Z_{ab}(s,x|A/\mu^2)$ for $(s,x)\in 
J_0\times I_u$, where $J_0$ and $I_u$ are respectively neighborhood
of $s=0$ in $\C$ and $u\in {\cal M}$.
Indeed, first of all,  no pole 
at $s=0$ arises in the functions $(s,x)\mapsto Z_{ab}(s,x|A/\mu^2)$
and in their $x$  derivatives. This is because, considering (\ref{Zeta}) and 
 (c2) of {\bf Theorem 2.1},
one notices that if any pole appears in the various $\zeta^{(\alpha,\beta)}$
 functions used building up  $Z_{ab}$
it has to be a simple pole. Anyhow, the factor $s$ makes the global
functions $Z_{ab}$ regular at $s=0$. 
Using recursively (c2) of {\bf Theorem 2.1}
one has that each function  $x\mapsto Z_{ab}(s,x|A/\mu^2)$
is $C^\infty$ in a neighborhood of  $u$ for any fixed $u\in {\cal M}$
and $s=0$. More generally, this results
holds for $s$ fixed in  neighborhood of $0$ because,
by (c1) of {\bf Theorem 2.1}, one has that  
no pole can arise in a open disk centered in $s=0$ with radius
$\rho =1/2$.  The functions $Z_{ab}$ and all their
$x$ derivatives are  also $s$-analytic for $x$ fixed in a neighborhood of
$0$. Then, we can conclude that
  any function $(s,x) \mapsto Z_{ab}(s,x|A/\mu^2)$
 is $C^\infty$ in a neighborhood of $(0,u)$  for any fixed $u\in {\cal M}$. 
The $C^\infty$
 smoothness of the stress tensor then follows trivially from (\ref{ztensor})
directly. 

The property (b) can be proved as follows. From the point (c2) of 
{\bf Theorem 2.1} and taking account of (a) of {\bf Theorem 2.3} and
the definition (\ref{ztensor}),
we have that (b) holds true if $\nabla^a Z_{ab}(s,x|A/\mu^2) = 
-Z(s,x|A/\mu^2) \nabla_b V'(x)$
 for the
considered point $x$ and  $s\in \C$ away from the poles, the function 
$Z$
in the right hand side that of the field fluctuations 
(see {\bf Definition 2.7} in \cite{m1}).  By the 
theorem of the uniqueness of the analytic continuation, if one is able to
 prove such an identity 
for $Re$ $s$ sufficiently large
 this assures also 
the validity of $\nabla^a Z_{ab}(s,x|A/\mu^2) 
= -Z(s,x|A/\mu^2) \nabla_b V'(x)$ 
everywhere
in the variable $s$. Therefore, 
let us prove that there is a $M>0$ such that 
$\nabla^a Z_{ab}(s,x|A/\mu^2) = -Z(s,x|A/\mu^2) \nabla_b V'(x)$ 
for $Re$ $s>M$ and this will be 
enough to prove the point (b). To get this goal
we represent $\nabla^a Z_{ab}(s,x|A/\mu^2)$ employing (\ref{sotto})
for each function $Z_{ab}$. Then we make recursive use of (\ref{piu'uno})
of {\bf Theorem 2.1} and obtain $\nabla^a Z_{ab}(s,x|A/\mu^2)$ 
written as a linear combination of functions
 $\zeta^{(\alpha,\beta)}(s+1,x|A/\mu^2)$. Finally we can expand all these
functions in series of the form  (\ref{serie'}) of {\bf Theorem 2.2},
provided $Re$ $s>M$ for an opportune $M>0$. Taking account of the comment
(4) after {\bf Definition 2.3}, the explicit expression 
of the final series of $\nabla^a Z_{ab}(s,x|A/\mu^2)$ reads, for $Re$ $s>M$
\begin{eqnarray}
\nabla^a Z_{ab}(s,x|A/\mu^2) =
s  {\sum_{j\in \N}}' 
\frac{2}{\lambda_j}
\left(\frac{\lambda_j}{\mu^2}\right)^{-s} 
\nabla^a \left\{  T_{ab}[\phi_j,\phi_j^*, {\bf g}](x) + \frac{\lambda_j 
g_{ab}(x)}{2}
 \phi_j(x) \phi_j^*(x)
\right\} 
 \label{Zeta2''}\:.
\end{eqnarray}
Finally, using the form of the $\zeta$ function of the
field fluctuatios given in \cite{m1}, 
one has to prove  that for any $x\in {\cal M}$
\begin{eqnarray}
\nabla^a \left\{  T_{ab}[\phi_j,\phi_j^*, {\bf g}](x) + \frac{\lambda_j 
g_{ab}(x)}{2}
 \phi_j(x) \phi_j^*(x)
\right\} = -\frac{1}{2}  \phi_j(x) \phi_j^*(x) \nabla_b V'(x) \label{temp}\:.
\end{eqnarray}
This is nothing but the generalized ``conservation law'' of the stress tensor
for the action 
\begin{eqnarray}
S_j[\phi,\phi^*] = \frac{1}{2}\int_{\cal M}
\left[\phi A'[{\bf g}] \phi^* - \lambda_j \phi \phi^*   \right] d\mu_g
\end{eqnarray}
Indeed, (\ref{temp}) holds when the field $\phi$ satisfies the motion 
equations for the action above
$A'\phi = \lambda_j \phi$. This is satisfied by the $C^{\infty}$
eigenfunctions of $A$ $\phi_j$ with eigenvalue $\lambda_j$. Therefore
(\ref{temp}) holds true and (b) is proven.

Concerning the point (c),  (\ref{t2}) 
with $t_{ab}(x|A) = Z_{ab}(0,x|A)/2$ arises as a direct 
consequence of the definition (\ref{ztensor}), noticing   that, from 
{\bf Theorem 2.3}, $Z(s,x|A/\mu^2)$ 
is analytic at $s=0$ and, from  (\ref{Zeta}),
(\ref{ztensor}) can be written down also
\begin{eqnarray}
\langle T_{ab}(x|A) \rangle_{\mu^2} = \frac{1}{2} \frac{d\:}{ds}|_{s=0}
Z_{ab}(s,x|A) 
+ \frac{1}{2} Z_{ab}(0,x|A) \ln\mu^2   \label{ztensor'}\:,
\end{eqnarray}
Similarly, from {\bf Definition 2.3}, one 
sees also that  $Z(s,x|A/\mu^2)/2$ evaluated at $s=0$ takes contribution
only from the possible pole at $s=1$ 
of the function $s\mapsto \zeta_{ab}(s,x|A/\mu^2)$
 (the remaining simple 
$\zeta$  function is regular for $s=0$) and coincides with
the value of residue of 
the pole of this function at $s=0$. When $D$ is odd, no pole
of $s\mapsto \zeta_{ab}(s,x|A/\mu^2)$  arises at $s=1$ because of (c)
of {\bf Theorem 2.1}, this is the reason 
for the $\delta_D$ in the right hand side of (\ref{twald}). The 
form (\ref{twald}) of $t_{ab}$
assures that it is built up as a liner combination of product of the metric,
curvature tensors, $V'(x)$ and their covariant derivatives, everything
evaluated at the same point $x$. 
The property (\ref{aggiunta}) is  consequence 
of $\nabla^a Z_{ab}(s,x|A/\mu^2) = -Z(s,x|A/\mu^2) \nabla_b V'(x)$ 
 proven during the proof of (b), employing the pole structure of 
the function $Z(s,x)$ given in \cite{m1}.
Notice also that $t_{ab}$ is symmetric by construction.
 Therefore, we have to prove
(\ref{twald}) and this conclude the proof of (c).
It is sufficient to show that
\begin{eqnarray}
\lim_{s\rightarrow0}
s \zeta_{ab}(s+1,x |A) &=& \frac{\delta_D}{(4\pi)^{D/2}}
\left\{  a_{D/2-1,(ab)}(x,x|A)
+ \frac{g_{ab}(x)}{2} a_{D/2}(x,x|A)  \right. \nonumber \\
&+& \left. \left[ \left(\xi -\frac{1}{4}\right)g_{ab}(x) \Delta + \xi 
R_{ab}(x) -\xi \nabla_a \nabla_b  \right] a_{D/2-1}(x,x|A)\right\} 
\:.\nonumber
\end{eqnarray}
From (a) of {\bf Theorem 2.3} this is equivalent to 
\begin{eqnarray}
 & & \lim_{s\rightarrow0} s \frac{1}{2}\left[\zeta^{(1_a,1_b)}(s+1,x |A) 
+\zeta^{(1_b,1_a)}(s+1,x |A) \right]  \nonumber \\
&+&  \lim_{s\rightarrow0} s\left[\left(\xi - \frac{1}{4}\right) 
 g_{ab}(x) \Delta + \xi R_{ab}(x) - \xi \nabla_a \nabla_b \right] 
\zeta(s+1,x|A/\mu^2)
\nonumber\\
 &=& \frac{\delta_D}{(4\pi)^{D/2}}
 \left\{a_{D/2-1,(ab)}(x,x|A) 
+ \frac{g_{ab}(x)}{2} a_{D/2}(x,x|A)\right.\nonumber \\
&+& \left. \left[ \left(\xi -\frac{1}{4}\right)g_{ab}(x)
 \Delta + \xi R_{ab}(x) -\xi
\nabla_a \nabla_b  \right] a_{D/2-1}(x,x|A)\right\} \label{M}\:.
\end{eqnarray}
The proof of this identity is very straightforward so we sketch its way only.
 Using {\bf Theorem 2.1},
it is sufficient
to consider the decomposition for large  $Re$ $s$  (and a similar
decomposition interchanging $a$ with $b$)
\begin{eqnarray} 
 s \zeta^{(1_a,1_b)}(s+1,x,y|A/\mu^2)|_{x=y} &=& \nonumber 
\end{eqnarray}
\begin{eqnarray}
& & \frac{s}{\Gamma(s+1)}
\int_0^{+\infty} dt\: t^{s}
D_x^{1_a} D_y^{1_b} \left[ 
K(t,x,y|A) - P_0(x,y|A) \right]|_{x=y} \nonumber \\
&=& \frac{s}{\Gamma(s+1)} \int_{0}^{\mu_{0}^{-2}}
 D_x^{1_a} D_y^{1_b} \{  \ldots \}|_{x=y} + \frac{s}{\Gamma(s+1)}
\int_{\mu_{0}^{-2}}^{+\infty}
D_x^{1_a} D_y^{1_b} \{  \ldots  \}|_{x=y}\:.  \label{Sbreak}
\end{eqnarray}
Then, we can expand the integrand the first integral in the second line of
(\ref{Sbreak}) using (\ref{first}) of 
{\bf Lemma 2.1} and we can continue both integral in the second line of 
(\ref{Sbreak}) as far as $s=0$.
A direct computation proves that, because of the 
presence of the factor $s$, only the first integral in  
(\ref{Sbreak}), expanded as said above, gives contribution.
The contribution arises from the terms of the heat-kernel expansion
which, once integrated in $t$ (taking account of the factor $t^{s}$
in the integrand), have a pole for $s=0$. This pole is canceled out
by the factor $s$ giving a finite result. The terms which have no pole 
at $s=0$ vanish due to the factor $s$, in the limit $s\rightarrow 0$. 
A similar procedure can be employed concerning the second limit in (\ref{M}).
In performing calculations, it is worth to remind that 
$\nabla_{(x)a} \sigma(x,y)$ and $\nabla_{(y)b} \sigma(x,y)$ {\em vanish}
in the limit $x\rightarrow y$, and furthermore 
$\nabla_{(x)a}\nabla_{(y)b} \sigma(x,y)  |_{x=y} = - g_{ab}(y)$.  
Summing all contributions one obtains (\ref{twald}).

Concerning the point (d), the proof is dealt with as follows.
Starting from (\ref{tab'}) one finds
\begin{eqnarray}
g^{ab} T_{ab}[\phi_j,\phi^*_j,{\bf g}] =
\nabla_c\phi_j\nabla^c \phi_j^* + \left\{\xi R + 
\left[\xi \left(D-1 \right) -\frac{D}{4} \right]
\Delta \right\} |\phi_j|^2 - \frac{D}{2} \lambda_j |\phi_j| \:.\nonumber
\end{eqnarray}
Then, employing the identities $2 \nabla_c \phi^*\nabla^c \phi =
\Delta |\phi|^2 - \phi \Delta \phi^* -\phi^* \Delta \phi$ and
 $(-\Delta + \xi R + m^2 + V') \phi_j= \lambda_j \phi_j$ we have also
\begin{eqnarray}
g^{ab} T_{ab}[\phi_j,\phi^*_j,{\bf g}] =
 \left[\xi \left( D-1\right) -\frac{D-2}{4}
 \right] \Delta |\phi_j|^2
-(m^2 +V')|\phi|^2
+ \frac{2-D}{2} \lambda_j |\phi_j| \nonumber\:.
\end{eqnarray}
Since $\xi_D = (D-2)/[4(D-1)]$, we have finally
\begin{eqnarray}
g^{ab} T_{ab}[\phi_j,\phi^*_j,{\bf g}] =
 \left[\frac{\xi_D-\xi}{4\xi_D-1}\Delta -m^2 -V'  \right]
  |\phi_j|^2
+ \frac{2-D}{2} \lambda_j |\phi_j| \nonumber\:.
\end{eqnarray}
From {\bf Definition 2.3}, this entails that, for $Re$ $s$ sufficiently large
\begin{eqnarray}
g^{ab}Z_{ab}(s,x|\mu^2) = \frac{2}{\mu^2} 
 \left[\frac{\xi_D-\xi}{4\xi_D-1}\Delta -m^2 -V'  \right]
s\zeta(s+1,x |A/\mu^2) + 2s\zeta(s,x|A/\mu^2) \label{tend}\:.
\end{eqnarray}
The function $(s,x) \mapsto s\zeta(s+1,x |A/\mu^2)$ is
 $C^\infty$ in a neighborhood of $(0,u)$ for any $u\in {\cal M}$
the proof is similar to that given in (b) above for 
 $(s,x) \mapsto Z_{ab}(s,x|\mu^2)$. 
Finally, employing {\bf Definition 2.4} taking also 
account of {\bf Definition 2.7} 
in \cite{m1} and (34) in {\bf Theorem 2.2} in \cite{m1}
 (i.e. (\ref{sto}) below),
one finds (\ref{t3}).$\Box$\\

\noindent {\em Comments}.\\
{\bf (1)}  Concerning the point (b) which generalizes
the classical law (\ref{conservation}), we stress that this result is strongly 
untrivial. We have not put this result somewhere ``by hand'' 
in the definitions and hypotheses we have employed. 
Notice also that, in the case $V'\equiv 0$,
  the tensor $T_{ab}[\phi_j,\phi^*_j, {\bf g}]$
we have used in the definitions is not conserved. Nevertheless, the final
stress tensor is conserved. This should means that the local 
$\zeta$ function approach is a quite  deep approach.\\
{\bf (2)}  Concerning the point (c), we notice that this result is in agreement
with Wald's axioms \cite{wald94} and, on a purely 
mathematical ground, it reduces
the ambiguity allowed by Wald's theorem. Indeed, Wald's theorem involves
at least two arbitrary terms dependent on two free parameters. Recently
it has been proven that in the case of massive field which are not
conformally coupled
such an ambiguity should be much larger \cite{tifl}.
The point (d) proves that the corresponding ambiguity related to the 
field fluctuations is consistent with that which arises from the stress
tensor.
Assuming the $\zeta$ function procedure the only ambiguity remaining is 
just that related to the initial arbitrary mass scale $\mu$. On the other
hand there is no  physical evidence that the $\zeta$-function
procedure is the physically correct one and thus one cannot conclude that 
this method gets rid of the ambiguity pointed out by Wald et al.  \\
{\bf (3)}  Concerning the point (d), we notice that, in the case 
$\xi =\xi_D$ and $V' \equiv m^2 = 0$ the usual 
conformal anomaly \cite{wa,ha} arises provided 
$D$ is even and the kernel of $A$ is trivial. Anyhow, 
in the case $Ker$ $A$ is 
untrivial, 
the trace anomaly takes a contribution from the null modes also when $D$
is odd.
In any cases, for the anomalous term, it holds ((34) in {\bf Theorem 2.2}
in \cite{m1})
\begin{eqnarray}
\delta_D \frac{a_{D/2}(x,x|A)}{(4\pi)^{D/2}} - P_{0}(x,x|A)
= \zeta(0,x|A/\mu^2)  \label{sto}
\end{eqnarray}
also for $\xi \neq \xi_D$.\\

Let us consider some issues related to the physical interpretations of the
 theory. \\
Suppose  $S_1$ acts as a globally  one-parameter isometry group  on 
the Riemannian manifold  ${\cal M}$
giving rise to closed orbits with period $\beta>0$. Suppose also that 
there exist a $D-1$ embedded submanifold $\Sigma$
which intersects each orbit just once and 
is orthogonal to the Killing vector field of the isometry group $K$ (notice
 that any  submanifold $\Sigma_\tau$, obtained by the action on 
 $\Sigma$ of the isometry group on the points of $\Sigma$, remains orthogonal
to the Killing vector field). In this case the Riemannian metric is
said {\em static}, the parameter of the group $\tau$ is said the 
{\em Euclidean time} of the manifold with period $\beta$ and the submanifold
$\Sigma$ is said the {\em Euclidean space} of the manifold.

As is well known, $\beta$ is interpreted as the
 ``statistical mechanics'' inverse temperature of the quantum state, 
anyway, it has no direct physical 
meaning because it can be changed by rescaling the normalization of the Killing
vector $K$ everywhere by a constant factor. 
The physical temperature which, in principle, may be measured by 
a thermometer is the local rescaling-invariant 
Tolman temperature $T_T := 1/[\sqrt{(K,K)} \beta] $.

Whenever ${\cal M}$ is static and  $\Sigma$ is endowed with a global 
coordinate system $(x^1,\cdots,x^{D-1})\equiv \vec{x}$, ${\cal M}$ is endowed
with a natural coordinate system $(\tau,\vec{x})$, $\tau \in (0,\beta)$
$\vec{x}\in \Sigma -{\cal F}$, where ${\cal F}$ is the set of the fix points
of the group (which, anyhow,  may be empty).
 This coordinate system is obtained by the evolution 
of the coordinates on $\Sigma$ along the orbits of the isometry group 
and  is almost
global in the sense that is defined everywhere in ${\cal M}$ except for
the set of the (coincident) endpoints of each orbit at $\vec{x}$ constant
including the fix points of the group. This set has anyway 
negligible measure. Coordinates $(\tau,\vec{x})$ given above are said {\em
static coordinates}. Notice that, in these coordinates,
$\partial_\tau g_{ab} = 0$ and $g_{\tau \alpha } = 0$ for
 $\alpha =1,\cdots, D-1$ everywhere. Local static coordinates are defined
similarly.

The important result is that,  
under our general hypotheses, supposing also that ${\cal M}$ is static
and admits static coordinates $(\tau,\vec{x})$  and $V'$ does not depend on
 $\tau$ one has that
 the stress tensor {\em depends on $\vec{x}$ only} and satisfies everywhere 
\begin{eqnarray}
\langle T_{\tau \alpha}(\vec{x}|A) \rangle_{\mu^2} = 
\langle T_{\alpha \tau}(\vec{x}|A) \rangle_{\mu^2} =
0 \label{TTA} 
\end{eqnarray}
for $\alpha = 1,\cdots,D-1$.
The  remarkable point as far as the  physical ground is concerned, is that
this  result allows one  to look for analytic continuations towards
 Lorentzian metrics 
performing the analytical continuation $\tau \rightarrow it$ and without 
encountering  imaginary components of the
continued stress tensor.  
Notice that also $\langle \phi^2(x|A)\rangle_{\mu^2}$ and the effective 
Lagrangian ${\cal L}_{\scriptsize \mbox{eff}}(x|A)_{\mu^2}$ (see 
{\bf Definition 2.5} in \cite{m1})
do not depend on the temporal coordinate and,
moreover, all results contained in 
{\bf Theorem 2.4} hold true in the Lorentzian section of the manifold,
 considering the trivial analitic continuations 
of all the terms which appear in the thesis. One has:\\

\noindent {\bf Theorem 2.5.} {\em Within our hypotheses on 
${\cal M}$ and $A$,  suppose ${\cal M}$ is static with Euclidean
time $\tau \in (0,\beta)$ ($\beta >0$) and $V'$ is invariant under
Euclidean time displacements.
In this case, for any $\mu^2>0$ and any point $x\in {\cal M}$, 
\begin{eqnarray}
\langle T_{ab}(x|A) \rangle_{\mu^2}
 K^{a}(x) \sigma^{b}(x) = 0 \label{TTS}	\:,
\end{eqnarray}
where $K$ is the Killing vector field  associated to the time $\tau$
and $\sigma(x)$ is any vector orthogonal to $K$ at $x$.
Furthermore, denoting  the Lie derivative along $K$ by ${\cal L}(K)$, 
it holds everywhere
on ${\cal M}$ 
\begin{eqnarray}
{\cal L}(K)_{c} \langle T^{ab}(x|A)\rangle_{\mu^2} &=& 0\:,  \label{lie1}\\
{\cal L}(K) \langle \phi^2(x|A)\rangle_{\mu^2} &=& 0\:, \label{lie2} \\
{\cal L}(K) {\cal L}_{\scriptsize \mbox{eff}}(x|A)_{\mu^2} &=& 0 
\label{lie3}\:.
\end{eqnarray}}
{\em 
Finally, all the results of {\bf Theorem 2.4} hold in the Lorentzian section
provided one considers the Lorentzian-time-continued quantities in place of
the corresponding Riemannian ones everywhere.}\\

\noindent {\em Scketch of Proof.}
In the given hypotheses and fixed $x\in {\cal M}$ ($x$ different
 from any fixed point
in such a case the thesis being trivial since $K(x) = 0$),
let us consider a generally local 
coordinate system $\vec{x}$ on the Euclidean space  $\Sigma$ around the 
intersection of the orbit passing from $x\in {\cal M}$, this induces 
a natural local coordinate system on ${\cal M}$, $(\tau,\vec{x})$
(where $\tau \in (0,\beta)$) 
which includes the same point $x$. In our hypotheses
(\ref{TTS}) is trivially equivalent to (\ref{TTA}) in the considered 
coordinate system.\\
Concerning  the form of the $\zeta$ function 
of the stress tensor given in {\bf Definition 2.3} taking account of 
(\ref{stresspratico'}),
since $g_{\tau\alpha}(x)=0$ and
$\partial_\tau g_{ab}(x) =0$, only the first line of 
(\ref{stresspratico'}) and the last term in the last line may produce 
the considered components of the stress tensor. 
Actually, the dependence from $\tau$ of the eigenfunctions $\phi_j$
 of the operator $A$, can be taken  of the 
form $e^{i \omega \tau}$ with $\omega \in \R$
 just because $\partial_\tau = K$
is a Killing field as we shall prove shortly. 
Then, the last term in the last line of
 (\ref{stresspratico'})
immediately vanishes concerning the considered components 
because the argument of the covariant derivatives (which commute 
on scalar fields) does not depend on $\tau$; furthermore,
taking account that $A$ is real and thus
$\phi_j$ and $\phi^{*}_j$ correspond to the same eigenvalue, one sees that
the contribution coming from the 
first line of (\ref{stresspratico'}) computed for $b\neq a=\tau$ and 
$a\neq b=\tau$
vanishes when one sum over $j$ to get the stress-tensor 
$\zeta$ function. The validity of (\ref{lie1}), (\ref{lie2}), (\ref{lie3}) 
 is also obvious working in local static coordinates where the Lie 
derivative reduces to the ordinary $\tau$ derivative and 
taking account of the imaginary exponential 
dependence form $\tau$ of the modes. In fact, this  dependence is canceled out
directly in the various $\zeta$ functions  due to the product of 
$\phi_j$ and $\phi^*_j$ (or corresponding derivatives) which appear in 
their definitions.

Let us finally 
prove that one can define the normalized orthogonal 
 eigenvectors of $A'$ (and thus $A$) 
in order to have the 
dependence from $\tau$ said above.
Reminding that each eigenfunction of $A$ is
a  $C^{\infty}({\cal M})$ function, 
 and working in the local coordinate system around the
 orbit of $x$
considered above where $g_{ab}$ does not depend on $\tau$,
one trivially has that
\begin{eqnarray}
A' \partial_\tau \phi_{jk_j} = \partial_\tau   A' 
 \phi_{jk_j} = \lambda_j \partial_\tau\phi_{jk_j} \label{ftl}\:,
\end{eqnarray}
where $\phi_{jk_j}$ is an eigenvector of $A$ with eigenvalue
$\lambda_j$.
This holds  in the considered coordinates and therefore, choosing
different local coordinate systems in $\Sigma$ and reasoning similarly,
the above identity can be proven to hold 
 {\em almost} everywhere in ${\cal M}$  provided 
$ \partial_\tau \phi_{jk_j}$ is interpreted as the 
$C^{\infty}({\cal M})$ scalar field
$(K, \nabla \phi_{jk_j})$.
Reminding that the
dimension of each eigenspace $d_j$ is finite ({\bf Theorem 1.1.} in 
\cite{m1}), it must be
\begin{eqnarray}
\partial_\tau \phi_{jk_j}(x) =
\sum_{l_j=1}^{d_j} 
c_{k_jl_j} 
\phi_{jl_j}(x) \label{ftl2}\:,
\end{eqnarray}
almost everywhere. 
Remind that locally 
$\partial_\tau g_{ab} =0$ and, since $g_{\tau\alpha}=0$,
 $g = (K,K) h$ where $h$ is the determinant
of the metric induced in $\Sigma$), one finds from (\ref{ftl2})
\begin{eqnarray}
c^*_{k_jh_j} + c_{h_jk_j}  &=&
\int_{{\cal M}} \partial_\tau \left\{ \phi^*_{jk_j}(x) \phi_{jh_j}(x) 
\right\} d\mu_g(x)\nonumber\\ 
&=& \int_{0}^{\beta}d\tau \frac{d}{dt}
\int_\Sigma
 \phi^*_{jk_j}(\tau,p) 
\phi_{jh_j}(\tau,p) (K(p),K(p))^{1/2}
  d\nu(p)  \nonumber
\:,
\end{eqnarray}
where $p$ is any point of the submanifold $\Sigma$ and
$\nu$ is its (finite) Riemannian measure induced there from the metric.
We have passed the derivative through the symbol of integration employing
Fubini's theorem and Lebesgue's dominate convergence theorem.
The right hand side of the identity above vanishes taking account 
that, for any fixed $p$ 
\begin{eqnarray}
\lim_{\tau\rightarrow 0^+} \phi(\tau,p)_{jk_j}  =
\lim_{\tau\rightarrow \beta^-} \phi(\tau,p)_{jk_j} \label{lim}  
\end{eqnarray}
because the orbits of the coordinate $\tau$ are closed and 
the functions are continuous in the whole manifold.
Therefore,
the matrix of the coefficients $c_{pq}$ is anti-hermitian.
Finally,
in the considered eigenspace, we can choose an orthogonal
 base of smooth normalized eigenfunctions
where the matrix above is represented by a diagonal matrix, the eigenvalues
being $i\omega_{jl_j}$, $\omega_{jl_j} \in \R$ and $l_j = 1,2,\cdots, d_j$.
In the new base one re-writes (\ref{ftl2}), in local coordinates,
\begin{eqnarray}
\partial_\tau \phi_{jk_j}(x) = i\omega_{j k_j}
\phi_{jk_j}(x) \label{ftl4}\:,
\end{eqnarray}
and this entails trivially, with $\omega_{j k_j} = 2\pi n_{j k_j}/\beta$, 
$n_{j k_j} \in \Z$
by (\ref{lim}),
\begin{eqnarray}
\phi_{jk_j}(x) = 
e^{i\omega_{jk_j}\tau} \varphi_{jk_j}(x^1,\cdots,x^{D-1})\:.
\end{eqnarray}
We leave to the reader the simple proof of the last statement 
of our theorem which can be carried out in local coordinates.  $\Box$\\

As a final remark notice that changes in the period $\beta$ of the 
the manifold which correspond to actual increases  of the proper length
of the orbits (and not to a simple rescaling of the normalization of the
Killing vector), in general produce {\em conical singularities}
in the fix points of the Lie group provided they exist. 
In such a case the manifold fails to be smooth and,
in general, the theorems proven in this work and in \cite{m1} may not hold.

\section*{IV.  The relation between the  $\zeta$ function and the 
point-splitting to renormalize the stress tensor. An improved formula
for the point-splitting procedure.}

Similarly to the previous work, we  prove here that, in our general 
hypotheses, a particular (improved) form of the  point-splitting
procedure can be considered as a 
consequence of the $\zeta$ function technique.

\subsection*{A. The point-splitting renormalization.}

Let us summarize the point-splitting approach to renormalize the
one-loop stress tensor \cite{bd,wald94,fu} in the Euclidean case.
 First of all, we want to
rewrite (\ref{stresspratico}) 
into a more convenient form. Employing the
motion equations $A'\phi \equiv 0$ one can rewrite the right hand side
of (\ref{stresspratico})  as
\begin{eqnarray}
T_{ab}[\phi,{\bf g}](x) &=&
(1-2\xi) \left[\nabla_a \phi(x)\nabla_b \phi(x) + 
\phi(x) \nabla_a \nabla_b \phi(x)   
\right] \nonumber\\
&+ & \left(2\xi -\frac{1}{2}\right) g_{ab}(x)
 \left[\nabla_c \phi(x)\nabla^c 
\phi(x) + \phi(x) \Delta \phi(x)   \right] \nonumber\\
&+& \left[  \frac{g_{ab}(x)}{D} \phi(x) \Delta\phi(x)  - 
\phi(x) \nabla_a \nabla_b \phi(x)    \right]\nonumber \\
&-& \xi \left[ R_{ab}(x) -\frac{g_{ab}(x)}{D}R(x)  \right]\phi^2(x) \nonumber
\\
&-& \frac{V'(x)+m^2}{D} g_{ab}(x) \phi^2(x) \label{Tgrosso}\:.
\end{eqnarray}
Notice that the first two lines in the right hand side of (\ref{Tgrosso})
produce a vanishing trace in the case of $\xi = \xi_D$  
($ := (D-2)/[4(D-1)])$, the third and the 
fourth line have separately a vanishing trace not depending on $\xi$.
Finally, the trace of the last line is $-[V'(x) + m^2]\phi^2(x)$ trivially.
It is obvious that, in the case of conformal coupling ($\xi=\xi_D$,
$V'\equiv m^2 = 0$), the trace of the stress tensor vanishes.
Conversely, for $\xi\neq \xi_D$ one get also
\begin{eqnarray}
g^{ab}(x)T_{ab}(x) = \left(\frac{\xi_D-\xi}{4\xi_D-1}
\Delta - m^2 - V'(x) \right) \phi^2(x) \label{trclass}\:.
\end{eqnarray}
This is nothing but the {\em classical} 
version of (\ref{t3}). The most important
difference is the lack of the  trace anomaly term which is related to the 
last two terms in the right hand side of (\ref{t3}). 

The point-splitting procedure can be carried out employing the expression above
for the stress tensor (actually one expects that the same final result 
should arise starting from different but equivalent expressions of the stress 
tensor). The basic idea is very simple \cite{bd,wald94,fu,ch76,ch78}. 
One {\em defines}
the $a,b$ component of the one-loop renormalized stress tensor 
in the point $y$ as the result of the following limit
\begin{eqnarray}
\langle T_{ab}(y) \rangle :=
 \lim_{x\rightarrow y} {\cal D}_{ab}(x,y)
 \left\{\: \langle \phi(x)
\phi(y) \rangle -  H(x,y) \:\right\} \:, \label{psdef}
\end{eqnarray}
where the quantum average of the couple of fields is interpreted as the Green 
function of the field equation corresponding to the quantum state one
is considering, $H(x,y)$ is a Hadamard local fundamental solution 
\cite{wald94,fu,garabedian} 
which has just the task to remove the argument-coincidence divergences 
from the Green function {\em and from its derivatives}
and does not depend on the quantum state.
The operator ${\cal D}_{ab}(x,y)$ ``splits'' the point $y$ and it is written
down following (\ref{Tgrosso}), after an  opportune symmetrization of the 
arguments (again, the final result should not depend on this
symmetrization procedure),
\begin{eqnarray}
{\cal D}_{ab}(x,y) &=&
\frac{1-2\xi}{2} \left[ I_a^{a'} \nabla_{(x)a'} \nabla_{(y)b}  + 
I_b^{b'}\nabla_{(x)b'} \nabla_{(y)a}  + 
 \nabla_{(y)a} \nabla_{(y)b} + I_a^{a'} I_b^{b'} 
\nabla_{(x)a'} \nabla_{(x)b'}  \right] \nonumber\\
&+ &  \left(2\xi -\frac{1}{2}\right) \frac{g_{ab}(y)}{2}
\left[2 I^{c'}_{c} \nabla_{(x)c'} \nabla^{(y)c}  +  \Delta_x +\Delta_y    
\right] \nonumber\\
&+& \frac{1}{2}\left[  \frac{g_{ab}(y)}{D} \left( \Delta_x  +
\Delta_y\right)  - 
 \nabla_{(y)a} \nabla_{(y)b} - I_{a}^{a'}I_{b}^{b'}
\nabla_{(x)a'} \nabla_{(x)b'} \right]\nonumber \\
&+& \xi \left[ R_{ab}(y) -\frac{g_{ab}(y)}{D}R(y)  \right] \nonumber
\\
&-& \frac{V'(y)+m^2}{D} g_{ab}(y)  \label{operator}\:.
\end{eqnarray}
Above $I_{a}^{b'} = I_{(y)a}^{(x)b'}(y,x)$ 
is a generic component of the bitensor
of  parallel displacement from $y$ to $x$, so the (co)tangent space at 
the point $x$ is identified with the fixed (co)tangent space at the point $y$.

What one has to fix, in order to use (\ref{psdef}) for a particular quantum
state, is the Hadamard solution
$H$. It is known that, in the case  $D$ is even, this solution is not unique
\cite{wald94,fu} and is determined once one has fixed the term 
$w_{0}(x,y)$ (see Comment (2) of {\bf Theorem 2.6} in \cite{m1}). This term,
differently from the case of the renormalization of the field fluctuations,
 is not completely arbitrary. Indeed, it is possible to show
that there are terms $w_{0}$ producing a
 left hand side of (\ref{psdef}) which is {\em not} conserved
 \cite{wald94}. Moreover, the massless conformally coupled 
case, and more generally, the case $m=0$ and $V'\equiv 0$,
 involves some difficulties for the choice of $w_0$. For $m\neq 0$, 
it is possible to fix $w_{0}$ through the Schwinger-deWitt algorithm
\cite{bd}  obtaining a conserved renormalized stress tensor
\cite{bd,wald94}. This is not possible for  $m=0$ because
Schwinger-deWitt's algorithm becomes singular in that case. Anyhow, there is 
a further prescription due to Adler, Lieberman and Ng \cite{aln}
(see also \cite{wald78,wald94}) which seems to overcome this drawback:
this is the simplest choice $w_0(x,y) \equiv 0$. 
However, in the case of a massless conformally coupled field at least,
 as pointed out by  Wald
\cite{wald78}, another drawback arises:
the above prescription cannot produce a conserved stress tensor. 
Nevertheless, as proven in 
\cite{wald78}, in the case of a  (analytic in the cited reference)
 either Lorentzian 
or Riemannian manifold, it is still possible to add a finite term 
in the right hand side of (\ref{psdef}) which takes account of the failure
of the conservation law in order to have a conserved final left hand side.
This further term carries also 
a contribution to the trace of the final tensor 
which then 
fails to vanish and coincides with the well-known conformal anomaly.
In \cite{wald94}, it has been argued that such an improved procedure can be
generalized to any value of $m$ and $\xi$ getting 
\begin{eqnarray}
\langle T_{ab}(y) \rangle :=
 g_{ab}(y) Q(y) + \lim_{x\rightarrow y} {\cal D}_{ab}(x,y)
 \left\{\: \langle \phi(x)
\phi(y) \rangle -  H^{(0)}(x,y) \:\right\} \:, \label{psdef2}
\end{eqnarray}
where $H^{(0)}$ is the Hadamard solution determined by the choice
 $w_0 \equiv 0$
and $Q$ is a term fixed by imposing both the conservation of the left
hand side of (\ref{psdef2}) and the request that the renormalized stress
tensor vanishes in the Minkowski vacuum. Employing the local 
$\zeta$-function approach, we shall find out a point-splitting
procedure which, in the case of a compact manifold, generalizes Wald's one 
for a general operator  $-\Delta + V$ in $D>1$ dimensions in a Riemannian,
not necessarily analytic, manifold and gives an 
explicit expression for $Q$ automatically.

\subsection*{B. Local $\zeta$ function and point-splitting procedure. 
An improved
point-splitting prescription.}

In this part of the work we shall state a theorem concerning
the relation between the two considered techniques proving their substantial
equivalence within our general hypotheses.\\

\noindent {\bf Theorem 3.1} {\em Let us assume our general hypotheses on 
${\cal M}$ and $A'$ and suppose also $D>1$.} 

(a) {\em The renormalized stress tensor
$\langle T_{ab}(y|A)\rangle_{\mu^2}$ defined in {\bf Definition 2.4}
 can be also computed as the result of
a point-splitting procedure. Indeed one has, for any $\mu^2>0$
\begin{eqnarray}
\langle T_{ab}(y|A)\rangle_{\mu^2} &=& 
 \lim_{x \rightarrow y}  {\cal D}_{ab}(x,y)
\left\{\: G(x,y|A) - H_{\mu^2}(x,y)   \: \right\}\nonumber\\
& & + \frac{g_{ab}(y)}{D}\left(
\delta_D \frac{a_{D/2}(y,y|A)}{(4\pi)^{D/2}} - P_{0}(y,y|A)\right) 
 \label{fine}
\end{eqnarray}
where ${\cal D}_{ab}$ is defined in} (\ref{operator}){\em , 
$G(x,y|A) := \mu^{-2} \zeta(1,x,y|A/\mu^2)$ is the $\mu^2$ independent 
``Green function'' of $A$ defined in} \cite{m1}{\em , $P_0(y,y|A)$ is the 
$C^{\infty}$ integral kernel of the projector on the kernel of $A$ and
$H_{\mu^2}(x,y)$ is defined as (the summation appears for $D \geq 4$ only)
\begin{eqnarray}
H_{\mu^2}(x,y) 
&=& \sum_{j=0}^{D/2-2} (D/2-j-2)!\left(\frac{2}{\sigma}\right)^{D/2-j-1}
\frac{a_{j}(x,y|A)}{(4\pi)^{D/2}}  
- \frac{a_{D/2-1}(x,y|A)}{(4\pi)^{D/2}} \left( 2\gamma 
+ \ln \mu^2\right) \nonumber\\
& & -\frac{2a_{D/2-1}(x,y|A) 
- a_{D/2}(x,y|A) \sigma}{2(4\pi)^{D/2}} \ln \left(\frac{\sigma}{2}\right)
 \label{unop}
\end{eqnarray}
if $D$ is even, and (the summation appears for $D\geq 5$ only)
\begin{eqnarray}
H_{\mu^2}(x,y) &=& \sum_{j=0}^{(D-5)/2}
\frac{(D-2j-4)!! \sqrt{\pi}}{2^{(D-3)/2-j}}
\left(\frac{2}{\sigma}\right)^{D/2-j-1}
\frac{a_{j}(x,y|A)}{(4\pi)^{D/2}} \nonumber \\
 & & +   \frac{a_{(D-3)/2}(x,y|A)}{(4\pi)^{D/2}}
\sqrt{\frac{2\pi}{\sigma}} 
-  \frac{a_{(D-1)/2}(x,y|A)}{(4\pi)^{D/2}}
\sqrt{2\pi\sigma} 
\label{unop'}
\end{eqnarray}
if $D$ is odd.}

(b) {\em $H_{\mu^2}$ is a particular Hadamard local solution 
of the operator $A'$ truncated at the orders $L,M,N$,
indeed one has
\begin{eqnarray}
H_{\mu^2}(x,y) &=&  
\frac{\Theta_D}{(4\pi)^{D/2}(\sigma/2)^{D/2-1}} 
\sum_{j=0}^L u_j(x,y) \sigma^j(x,y) +
 \delta_D \left(\sum_{j=0}^{M} v_j(x,y)\sigma^j\right) \ln \left( 
\frac{\sigma}{2}\right)\nonumber\\
&+& \delta_D \sum_{j=0}^{N}
w_{j}(x,y) \sigma^j  \label{duep}
\end{eqnarray}
where $\delta_D= 0$ if $D$ is odd and $\delta_D =1$ if $D$ is even;
  $\Theta_D = 0$ for $D=2$ and $\Theta_D =1$ otherwise; and furthermore}\\
(1) {\em  $L = D/2 -2$, $M=1$ and $N=0$ for $D$ even, and $L= (D-1)/2$
when $D$ is odd.}\\
(2) {\em The coefficients $u_j$ and $v_j$ of the above Hadamard expansion are 
completely determined 
by fixing the value as $x\rightarrow y$ of the coefficient of the
leading  divergent term in order that that this expansion for 
$L,M,N \rightarrow +\infty $ defines
a  Green function formally. Using our conventions, this means: 
\begin{eqnarray}
 u_{0}(y,y) = \frac{4\pi^{D/2}}{D(D-2) \omega_D} \label{n1}\:,
\end{eqnarray}
 for
$D\geq 3$, $\omega_D$ being the volume of the unitary $D$-dimensional
disk, and
\begin{eqnarray}
 v_{0}(y,y) = \frac{1}{4\pi} \label{n2}
\end{eqnarray}
 for $D=2$.}\\
(3) {\em 
The coefficients $w_j$, when $D$ is even, are completely determined by posing
\begin{eqnarray}
w_{0}(x,y) := -\frac{a_{D/2-1}(x,y|A)}{(4\pi)^{D/2}}  
\left[2\gamma + \ln \mu^2 \right] \:. \label{scelta}
\end{eqnarray}}

\noindent {\em Proof.} See {\bf Appendix }. $\Box$\\

\noindent {\em Comments}\\
{\bf (1)}  Whenever $D$ is even, the logarithm in  (\ref{duep})
 contains a dimensional quantity.
At first sight, this may look like a mistake. Actually, this apparent drawback
means that the third summation in (\ref{duep}) has to contain 
terms proportionals to $\ln \mu^2$ which can be reabsorbed 
in the second summation transforming the argument of the logarithm
from $\sigma/2$ into the nondimensional one $\sigma \mu^2/2$. 
Indeed, the term in the right hand side of (\ref{scelta}) makes this job
concerning the term $v_0$ in (\ref{duep}). Since (\ref{duep}) 
is computed up to $M=1$, 
one may expect the presence of a corresponding term $w_1$ in the
last summation in (\ref{duep}). Actually, this term gives no contribution
to the stress tensor employing (\ref{fine}) and (\ref{operator}) 
as one can check directly, taking account that in any coordinate system
 around any $x\in {\cal M}$ (with an obvious meaning
of the notations) 
\begin{eqnarray}
I^{a'}_b(x,y)|_{x=y} = \delta_b^{a'}
\end{eqnarray}
and 
\begin{eqnarray}
\nabla_{(x)a}\nabla_{(x)b}\sigma(x,y)|_{x=y} &=& -\nabla_{(x)a}\nabla_{(y)b}
\sigma(x,y)|_{x=y}\nonumber \\
 = \nabla_{(y)a}\nabla_{(y)b}\sigma(x,y)|_{x=y} &=&
-\nabla_{(y)a}\nabla_{(x)b}\sigma(x,x')|_{x=y}
= g_{ab}(y) \label{sigmavarie}
\end{eqnarray}
In particular, one can check that each line of the  right hand side of 
(\ref{operator})  vanishes separately 
when it is evaluated for  the considered terms of 
the  Hadamard expansion and $x\rightarrow y$  
This is the reason because we have put  $N=0$ in (\ref{duep}) 
and we have omitted the corresponding term in (\ref{unop}).
Notice that, conversely, in the
usual version of point-splitting procedure \cite{bd,wald78,wald94} 
the term $w_1(x,y)$ is necessary. 
Similarly, the terms of order
$\sigma^n \ln \sigma $ with $n>1$ give no contribution to the stress
tensor and thus we have omitted them in (\ref{duep}). \\ 
{\bf (2)}  In the case $D$ is odd, the expansions
 (\ref{unop'}) and (\ref{duep}) do not consider
terms corresponding to $\sigma^{k +(1/2) }$ with $k=1,2, \cdots$. In fact
these terms give no contribution to the stress tensor via (\ref{fine})
and (\ref{operator}). Also in this case,  each line of right 
hand side of (\ref{operator}) gives a contribution which
 vanishes separately for $x\rightarrow y$.
Since (\ref{operator}) and (\ref{fine}) 
involve that the result does not depend from the coordinate 
system, one can check this fact working in  Riemannian
normal coordinates centered in $y$. \\
{\bf (3)}  The point-splitting procedure suggested in \cite{wald94} for $D=4$, 
differently 
from our procedure, requires $w_{0}\equiv 0$ rather than (\ref{scelta}).
Actually, the function $\sigma$ which appears in \cite{wald94} is defined as
two times our function $\sigma$. Therefore, taking account that
the argument of the logarithm in the second line of (\ref{unop}) is a quarter
of Wald's one, Wald's prescription corresponds to take
$w_0(x,y) - v_{0}(x,y) \ln 4 = 0$ 
in our case. Actually, as clarified 
in \cite{wald78}, the logarithm argument which appear in Wald's prescription
has to be understood as $\ln \left(\sigma/u^2 \right)$, where 
$u$ is the unit of length employed. In our formalism, this correspond, 
in particular, to
perform the changes $\ln (\sigma /2) \rightarrow \ln [\sigma/(2u^2)]$
and $\ln \mu^2 \rightarrow \ln (\mu^2 u^2)$ in (\ref{unop}).
(\ref{unop}) with the changes above
 entails that Wald's prescription, namely  
$w_0(x,y) - v_{0}(x,y) \ln 4 = 0$, 
is satisfied provided one fixes a $\mu^2$ such that 
$2\gamma + \ln ( \mu^2 u^2/4) = 0 $ namely
$\mu = 2 e^{-\gamma}/u$.\\
This proves that, under our hypotheses, our prescription generalize
Wald's one when the latter is understood in the Euclidean
 section of the manifold. Moreover, 
our prescription, differently from 
\cite{wald94},   gives explicitly the form of
the Hadamard local function to subtract to the Green function in the
general case as well as an explicit expression for 
the term $Q$ in (\ref{psdef2}), in terms of heat-kernel coefficients
and, trivially,
 for any choice of the value of $\mu^2$.\\
{\bf (4)}  Rescaling the parameter $\mu^2$, the expression of the final 
stress tensor changes by taking a term $(\ln \alpha) 
t_{ab}(y)$. We know the explicit form of
such a term, indeed, it must be that given
 in the point (c) of {\bf Theorem 2.4}.
Notice also that the obtained point-splitting method, 
also concerning the rescaling of $\mu^2$ agrees with 
the corresponding
point-splitting procedure for computing the field fluctuations given in
\cite{m1}. For example, the point (d) of {\bf Theorem 2.4} holds, provided
both sides are renormalized with the point-splitting procedures above and
the same value of $\mu^2$ is fixed.\\
{\bf (5)}  The point-splitting procedure we have found out uses the heat-kernel
expansion  in {\bf Theorem 1.3} of \cite{m1} 
and nothing further. This expansion can be built up also
either in noncompact manifolds or manifolds containing boundary, essentially
because it is based upon local considerations (see discussion in
\cite{wald78} concerning Schwinger-DeWitt's expansion). Therefore,
it is natural to expect that the obtained procedure, not depending
on the $\zeta$-function approach, may work in the general case (namely,
it should produce a symmetric conserved stress tensor with the known
 properties of the trace also in noncompact or containing boundary,
 manifolds), provided
the Green function of the considered quantum state has the Hadamard
behaviour.\\
{\bf (6)} As a final comment, let us check on the found point-splitting 
method in the Euclidean section of Minkowski spacetime which is out of
our general hypotheses, without referring to the
$\zeta$-function approach. In this case, for $A = -\Delta + m^2$, one has
that the heat kernel referred to globally flat coordinates reads
\begin{eqnarray}
K(t,x,y|A) = \frac{e^{-\sigma/2t}}{(4\pi)^2 t^2} e^{-m^2t} \label{hk}
\end{eqnarray}
and thus,  supposing $m^2>0$,
\begin{eqnarray}
a_{j}(x,y|A) = \frac{(-1)^j m^{2j}}{j!} \label{coe}
\end{eqnarray}
As is well known, 
the (Euclidean) Green function of Minkowski vacuum can be computed directly 
\begin{eqnarray}
G(x,y|A) &=& \int_0^{+\infty} K(t,x,y) dt = \frac{2m}{(4\pi)^2 \sqrt{\sigma/2}}
K_1\left(2\sqrt{\frac{\sigma m^2}{2}}\right) \:.
\end{eqnarray}
Expanding $K_1$ in powers and logarithms of $\sigma$ one get
\begin{eqnarray}
G(x,y|A) &=& \frac{2}{(4\pi)^2 \sigma} + \frac{1}{(4\pi)^2}\left\{ 
m^2 + \frac{m^4}{4} \sigma + \sigma^2 f(\sigma) \right\} \ln \left(
\frac{\sigma}{2}\right) \nonumber \\
& &+ \frac{m^2}{(4\pi)^2} \left( 2\gamma -1 + \ln m^2  \right) +
\sigma g(\sigma)\:,
\end{eqnarray}
where $f$ and $g$ are smooth bounded functions.
Then, employing (\ref{fine}), (\ref{duep}) and (\ref{coe}), it is a trivial
 task to prove that, {\em provided} the  choice $\mu = m e^{-3/4}$
is taken  in (\ref{scelta}), one gets
$<T_{ab}(y)> \equiv 0$ as it is expected. In particular one finds also
$Q(y) \equiv  m^4/ (128 \pi^2)$ for the coefficient of $g_{ab}(y)$
in the last line of 
(\ref{fine}).  For a general value of $\mu^2$,
 the computation of the stress-tensor trace via the formula in
(d) in {\bf Theorem 2.4} reproduces the correct Coleman-Weinberg results
\cite{cw} for the field fluctuations still obtained by the local 
$\zeta$ function approach \cite{dm} 
as well as by using the point-splitting formula
given in {\bf Theorem 2.6} in \cite{m1}. \\
The case $m=0$ is much more trivial. In this case the heat kernel is given by
 (\ref{hk}) with $m=0$, and thus only $a_{0}(x,y) \equiv 1$ survives
in the heat-kernel  expansion. In this case, $A$ is not positive defined
but positive only, the manifold is not compact and the Minkowski vacuum 
Green function can be still computed integrating the heat kernel despite 
the local $\zeta$ function does not exist. 
Moreover Green function coincides with  the
Hadamard local solution $2/[(4\pi)^2 \sigma]$, furthermore $Q(y) \equiv 0$,
 and thus our procedure 
gives a vanishing stress tensor as well.

\section*{V. Summary and outlooks.}

In this paper we have concluded the rigorous analysis started
in \cite{m1}, concerning the mathematical foundation of the theory of the
local $\zeta$-function renormalization of the one-loop stress tensor introduced
in \cite{moa}.
The other important point developed herein has been the 
relation between the local $\zeta$-function approach and the 
(Euclidean) point-splitting
 procedure.

Concerning the first point, we have proven that 
the $\zeta$-function theory of the stress tensor can be rigorously 
defined at least in closed manifold giving results which agree with
and generalize 
previous results concerning the $\zeta$-function renormalization
of the field fluctuations \cite{m1}. 
On the mathematical ground, we have also proven
a few of new theorems about the smoothness of the heat-kernel expansions.

Concerning the second proposed goal, we have found out that
the two methods ($\zeta$-function and point-splitting)
agree essentially, provided a particular form of  point-splitting procedure
is employed. Within the hypotheses of a Riemannian
compact $C^{\infty}$ manifold, 
this point-splitting procedure is a natural generalization
(in any $D>1$ and for a larger class of Euclidean motion operators)
 of Wald's improved procedure presented in \cite{wald78} and also discussed
in \cite{wald94} defined in a
Lorentzian   manifold (but the same arguments
employed can by trivially extended to Riemannian manifolds). Our procedure 
gives also explicitly the form of the various terms which are employed
in the point-splitting procedure in terms of the heat-kernel expansion.

In our opinion, the found point-splitting procedure should work also
without the employed hypotheses and independently from the $\zeta$-function
procedure. We have anyhow checked this conjecture in the Euclidean 
Minkowski spacetime 
proving that it holds true as expected either in the case $m=0$ or $m>0$.
Moreover, the obtained results concerning the point splitting procedure 
should be trivially generalized for static Lorentzian manifolds at least.

\section*{Appendix A. Proof of some lemmata and theorems.}

\noindent {\bf Proof of } {\bf Lemma 2.1}.
Let us consider the form of the heat kernel as it was 
built up in \cite{ch} Section 4
Chapter VI.  This construction holds also in the case of an operator 
$A' :=-\Delta+V$ and not only $A':= -\Delta$ as pointed out in the previous
 work
\cite{m1}. In our notations, one has by 
(45) in Section 4 of Chapter VI of 
\cite{ch} 
\begin{eqnarray}
F_N(t,x,y) = \frac{e^{-\sigma(x,y)/2t}}{(4\pi t)^{D/2}} \chi(\sigma(x,y))
\sum_{j=0}^{N}
a_j(x,y|A)t^j \:, \label{parametrics}
\end{eqnarray}
$N>D/2+2$ is a fixed integer.  \\
(Actually, the equation (45) in Section 4 of Chapter VI of 
\cite{ch} is missprinted in \cite{ch} because of
 the unnecessary presence of the
 operator $L_x$ in the right hand side 
of the first line of (45) in Section 4 of Chapter VI of \cite{ch}. 
Since the absence of this operator in the correct formula, we
 cannot get the second
line of (45) in a direct way.
 In fact, in \cite{m1} we have  used  a different
[but equivalent 
in the practice]
 form of the remaining of the heat-kernel expansion with respect to 
that which appears in (45). 
Some other parts of Section 4 of Chapter VI in \cite{ch}
 contain several other 
missprints like the requirement $F\in C^0(M\times M\times [0+\infty))$ 
in {\bf Lemma 2} which has to be corrected into 
$F\in C^1(M\times M\times [0+\infty))$.)
\begin{eqnarray}
K(t,x,y|A) = F_N(t,x,y) + (F_N * F)(t,x,y) \label{sopra}
\end{eqnarray}
where $F_N$ is the $C^{\infty}((0,+\infty)\times {\cal M}\times{\cal M})$
 parametrix defined in \cite{m1}.)\\
The remaining proportional to 
$O_{\eta,N}$ in (20) of 
{\bf Lemma 2.1} of \cite{m1} is therefore $(F_N * F)(t,x,y)$.\\
We  remind the reader that $\sigma(x,y)$ is one half the squared geodesical
distance ($d(x,y)$) from $x$ to $y$ and defines an everywhere continuous 
function  in  ${\cal M}\times{\cal M}$ which is also 
 $C^\infty$ in the set of the points
$x$,$y$ such that $d(x,y)<r$. $\chi(u)$ is a nonnegative 
$C^{\infty}([0,+\infty))$ function which takes the constant value $1$ for 
$|u|<r^2/16$ and vanishes for $|u|\geq r^2/4$, $r$ being the injectivity 
radius of the manifold. 
The convolution $*$ has been defined  in Section 4 of Chapter VI of \cite{ch}
\begin{eqnarray}
(G*H)(t,x,y):= \int_0^t d\tau \int_{\cal M} d\mu_g(z)
 G(\tau,x,z) H(t-\tau,z,y) \:,
\end{eqnarray}
whenever the right hand side makes sense.\\
Finally, the function $F$ which appears in (\ref{sopra})
is defined by a uniformly convergent series in $[0, T]\times{\cal M}
\times {\cal M}$ for any $T>0$
 (see (43) in Section 4 Chapter VI of \cite{ch}).
\begin{eqnarray}
F(t,x,y) := \sum_{l=1}^{\infty} [(A'_x-\partial/\partial t)F_N]^{*l}(t,x,y)\:.
\end{eqnarray}
($B^{*l}$ means $B*B*\cdots*B$ $l$ times.). This function belongs to
$C^L([0,+\infty)\times {\cal M}\times{\cal M})$ provided $M>D/2 +2L$
(see \cite{ch}).  (\ref{sopra}) satisfies
the heat-kernel equation provided $F$ is $C^{1}$ in all variables, namely 
$N> D/2 +2$. \\
Let us consider (\ref{sopra}). The remaining of the ``asymptotic'' expansion
 of
the heat kernel computed up to the coefficient  $a_N(x,y|A)$ ($N>D/2+2$)
is just the second term in the right hand side. It can be 
explicitly written down (see \cite{ch})
\begin{eqnarray}
(F_N * F)(t,x,y) & =& \int_0^{t} 
\:d\tau  
\tau^{-D/2} (t-\tau)^{N-D/2} \int_{\cal M} 
d\mu_g(z)\:  {\cal F}_N(\tau,x,z) {\cal F}(t-\tau,z,y)\times
 \nonumber\\
& & \times e^{-\sigma(x,z)/2\tau}e^{-\sigma(z,y)/2(t-\tau)}\:. \label{resto}
\end{eqnarray}
${\cal F}_N(t,x,z)$ defines a function which 
 belongs to
 $C^{\infty}([0,+\infty)\times {\cal M}\times {\cal M})$ and 
vanishes smoothly 
whenever the geodesical distance between $x$ and $z$ is sufficiently
large, i.e. $d(x,z)\geq r/2$,
 due to the presence of the function $\chi$ in the expression
of the parametrics (\ref{parametrics}).
$ {\cal F}$ defines an everywhere continuous function which 
 belongs also to
 $C^{L}$ provided $N>D/2 +2L$ and
 the geodesical distance between $y$ and $z$ is sufficiently
short, i.e. $d(y,z)< r$, and $t\in [0, +\infty)$.\\
Then let us pick out a point $u \in {\cal M}$.
 We can find a geodesically
spherical open 
neighborhood of $u$, $J_u$, with a geodesic radius $r_0<r/8$.
 By the 
definition of the function $\chi$, it holds $\chi(\sigma(x,y)) = 1$
whenever $x,y \in J_u$ and thus the coefficient $\chi$ can be omitted
in the heat-kernel expansion working with any coordinate system defined
in a neighborhood of $J_u$ (e.g. normal Riemannian coordinates). From now on
concerning the points $x$ and $y$ we shall work within such a coordinate 
system in the neighborhood $J_u$. Notice also that, by the triangular 
inequality $d(x,y) (= \sqrt{2\sigma(x,y)})<r/4$ whenever $x,y \in J_u$. \\
Now, let us suppose $N> D/2 +2|\alpha| + 2|\beta|$, this entails
$F \in C^{|\alpha|+|\beta|}$ and thus also ${\cal F} 
\in C^{|\alpha|+|\beta|}$ provided the distance of its arguments defined in 
the manifold   is less than $r$ and $t\in (0,+\infty)$.\\
 We can apply  operators $D_x^\alpha$ 
and  $D_y^\beta$ to both sides of (\ref{sopra}). 
The action of the derivatives
(\ref{sopra}) produces the first term in the right hand side of (\ref{first})
at least
(notice that $\chi \equiv 1$ in our hypotheses). Let us focus attention on the
action of the derivatives on the remaining in  
(\ref{sopra}). Our question concerns the  possibility
 to pass these under the integration
symbol in (\ref{resto}). 
The action
of the derivatives can be carried under integration symbols (obtaining
also $x,y$-continuous final function if the derivatives of the integrand
are continuous) provided, for any fixed choice of a couple of multindices
$\alpha,\beta$, the derivatives of the integrand
are  locally $x,y$-uniformly bounded by an integrable function (dependent
on the multindices in general). We shall see 
that this is the case after we have manipulated the integral opportunely.
Notice that the  
derivatives (with respect to the manifold variables) 
of the function ${\cal F}$ do exist 
because the second integral in the right hand side of (\ref{resto})
takes contribution only from the points $z$ such as both $d(y,z)<r$ and
$d(x,z)<r$ are fulfilled as required above.  
Indeed, it must be $d(x,z) < r/2 $ otherwise  
${\cal F}_N(\tau,x,z)$ smoothly  vanishes as pointed out
above, and, taking account of $d(x,y)< r/4$, the triangular inequality
entails also $d(y,z)\leq d(x,z)+ d(x,y) < r/2 +r/4 = 3r/4 $.\\
Now, let us fix a new open neighborhood of $u$, $I_u$,
 such 
that its closure is contained in $J_u$, and fix $T>0$.
Barring $\tau \mapsto \tau^{-D/2}$, 
all functions of $\tau,x,y,z$ and all their ($x,y,z$-)derivatives 
we shall consider are bounded in the compact 
$[0,T] \times \bar{I}_u \times \bar{I}_u \times{\cal M}$ where we are working
 because these are
continuous therein.
We can rearrange the expression (\ref{resto}) into
\begin{eqnarray}
(F_N * F)(t,x,y) & =& \int_0^{t} 
\:d\tau  
\tau^{-D/2} (t-\tau)^{N-D/2}  \int_{S^{D-1}} d\vec{v} 
\int_0^{+\infty} d\rho \rho^{D-1} J(x,\vec{v},\rho) \nonumber\\  
&\times&  e^{-\rho^2/2\tau} {\cal F}_M(\tau,x,z(x,\rho,\vec{v})) 
{\cal F}(t-\tau,z(x,\rho,\vec{v}),y)\nonumber\\
&\times& e^{-\sigma(z(x,\rho,\vec{v}),y)/2(t-\tau)}\:. \label{resto'}
\end{eqnarray}
where, to determine the position of $z$,
 we have employed a spherical system of normal coordinates
$\rho,\vec{v}$ centered in any $x$, $\rho$ is the distance of $z$ from $x$,
its range is maximized in the integrals above because the integrand
vanishes smoothly for  $\rho>r/2$, and thus all the functions contained in the 
integrand are well-defined within $\{\rho \in [0, +\infty)\}$.
 $\vec{v}$ is a unitary $D-1$ dimensional 
vector and $d\mu_g(z) = d\rho d\vec{v} \rho^{D-1} J(x,\vec{v},\rho)$,
$d\vec{v}$ is the natural measure in $S^{D-1}$. The function $J$ is continuous
and 
 bounded in $\bar{I}_u \times S^{D-1}\times \{\rho \in [0, r/2]\}$
 together with
all derivatives.\\
Then, we can change variables $\rho \mapsto \rho/\sqrt{\tau} =: \rho'$
obtaining
\begin{eqnarray}
(F_N * F)(t,x,y) & =& \int_0^{t} 
\:d\tau  (t-\tau)^{N-D/2} \int_{S^{D-1}} d\vec{v} 
\int_0^{+\infty} d\rho' \rho'^{D-1} J(x,\vec{v},\tau^{1/2}
\rho') \nonumber\\  
&\times&  e^{-\rho'^2/2} {\cal F}_M(\tau,x,z(x,\tau^{1/2}\rho',\vec{v})) 
{\cal F}(t-\tau,z(x,\tau^{1/2}\rho',\vec{v}),y)\nonumber\\
&\times& e^{-\sigma(z(x,\tau^{1/2}
\rho',\vec{v}),y)/2(t-\tau)}\:. \label{resto''}
\end{eqnarray}
The formal action  of the operators
$D^{\alpha}_x$ and $D^{\beta}_y$  
under the integration
produces a sum of continuous and bounded
functions (now the function $\tau \mapsto \tau^{-D/2}$ has disappeared
and the remaining functions and 
their $x,y,z$-derivatives  are bounded since they are
product of bounded functions). 
Also, it changes  $(t-\tau)^{N-D/2}$ into several terms of the form 
$(t-\tau)^{N-D/2 - L_i}$  (where each $L_i \leq |\alpha| + |\beta|$),
 because of the derivatives
of the second exponential.
 These function of $\tau$ are continuous and bounded
being $N > D/2 + |\alpha | + |\beta | \geq L_i$
 in our hypotheses. We can bound 
the absolute value of these functions
by  $ C e^{-\rho'^2/2} $, where $C$ is a sufficiently large constant.
This function is trivially integrable in the measure we are considering.
This $x,y,t$-uniform bound assures that, concerning the $x,y$-derivatives
of $F_N * F$,  one can interchange the
symbols of derivatives with those of integrals and also that the derivative
of $(t,x,y)\mapsto (F_N * F)(t,x,y)$ 
are continuous functions in $(0,+\infty)\times {\cal I}_u
\times {\cal I}_u$. \\
In order to finish this proof let us consider a finer estimate of 
$O^{(\alpha,\beta)}_{\eta,N}(x,y)$. We have
 the inequality \cite{ch}, for 
$\tau \in [0,t]$
\begin{eqnarray}
\frac{d^{2}(x,y)}{t} \leq \frac{d^{2}(x,z)}{\tau} +
\frac{d^{2}(z,y)}{t-\tau} \label{tri}\:,
\end{eqnarray}
and thus, picking out any 
$\eta\in(0,1)$ and posing $\delta := 1-\eta \in (0,1) $ we get
(notice that $t-\tau \geq 0$)
\begin{eqnarray}
e^{-\sigma(x,z)/2\tau}  e^{-\sigma(z,y)/2(t-\tau)}
&\leq& e^{-\eta\sigma(x,y)/2t}
\left(e^{-\delta \sigma(x,z)/2\tau}  e^{-\delta \sigma(z,y)/2(t-\tau)}\right)
\nonumber\\
&\leq& e^{-\eta\sigma(x,y)/2t} e^{-\delta \sigma(x,z)/2\tau} 
\:.
\end{eqnarray}
We can use this relation in the $x,y$-derivatives of  (\ref{resto''}) 
obtaining 
\begin{eqnarray}
|D_x^{\alpha}D_y^{\beta}(F_N * F)(t,x,y)| & \leq & \sum_i
 e^{-\eta\sigma(x,y)/2t}
\int_0^{t} \:d\tau  (t-\tau)^{N-D/2-L_i}
  \int_{S^{D-1}} d\vec{v} \nonumber\\
&\times& \int_0^{+\infty} d\rho' \rho'^{D-1}   e^{-\delta \rho'^2/2} C_i\:,
\end{eqnarray}
where the coefficients $C_i$ are upper bounds of the absolute values of
the continuous
  functions missed  in the integrand  and
$L_i \leq |\alpha|+|\beta|$. We can execute the integral in $\tau$
obtaining, for $0 < t\leq T $ and $x,y \in I_u$ (remind that
$N> D/2 +|\alpha| + |\beta| $)
\begin{eqnarray}
|D_x^{\alpha}D_y^{\beta}(F_N * F)(t,x,y)| &\leq&
 e^{-\eta\sigma(x,y)/2t}  \sum_i C'_{i,\delta}  
t^{N+1 - L_i - D/2}\nonumber\\
& \leq& \frac{C_{\delta}}{(4\pi)^{D/2}}  e^{-\eta\sigma(x,y)/2t} 
t^{N+1 - D/2 - |\alpha| - |\beta|}\:,
\end{eqnarray}
$C_\delta$ is a positive constant sufficiently large 
which depends on $T, \alpha, \beta$ in general.
This proves the remaining part of the thesis posing 
 $K_{\eta,N}^{(\alpha,\beta)} := T $ and
$M_{\eta,N}^{(\alpha,\beta)} := C_\delta $. 
Indeed, the remaining $O^{(\alpha,\beta)}_{\eta,N}$
we wanted to compute  coincides 
with $D_x^{\alpha}D_y^{\beta}(F_N * F)$ just up to  the  
factor $ (4\pi t)^{-D/2} t^{N -|\alpha|-|\beta|}\exp{(-\eta \sigma/2t)}$. 
$O^{(\alpha,\beta)}_{\eta,N}$ 
can be defined in $t=0$ as $O_{\eta,N}^{(\alpha,\beta)}(0;x,y) = 0$,
obtaining a continuous function in $[0,+\infty)\times I_u\times I_u$.
$\Box$\\

\noindent {\bf Proof of Lemma 2.2.}
Let us consider an eigenvector $\phi_j$ and fix $T\in (0,+\infty)$ and
consider a neighborhood of $u\in {\cal M}$, $J_u$ where a coordinate system
is defined. In the following,
 $x$ and $y$ are a points in a new neighborhood of $u$, $I_u$,
 such that its closure is contained 
in $J_u$. These points are  represented by the  coordinate system given
above and the derivative operators are referred to these coordinates. From 
{\bf Theorem 1.3} of \cite{m1},
it holds 
\begin{eqnarray}
e^{-T\lambda_j}\phi_j(x) 
= \int_{\cal M} d\mu_g(z)\:K(T,x,z|A)\phi_j(z) \:.
\end{eqnarray}
We can derive both sides of the equation above employing operators
$D_x^\alpha$. 
Since, for a fixed $T$ the derivatives of
$K$ are bounded ($(x,z) \mapsto K(T,x,z|A)$ is $C^{\infty}$  and $\bar{I}_u
\times {\cal M}$ is compact in our hypotheses \cite{m1}), we can pass
the derivative operator under the integral symbol obtaining
\begin{eqnarray}
|D_x^\alpha \phi_j(x)| = |e^{\lambda_j T}
\int_{\cal M} d\mu_g(z)\:
D_x^\alpha K(t,x,z|A) \phi_j(z)
| \leq e^{\lambda_j T} ||D_x^\alpha K(T,x,\:.\: |A)||_{L^2({\cal M},d\mu_g)}
\end{eqnarray}
where we have made use of the Cauchy-Schwarz inequality and we have
taken account of $||\phi_j|| =1$ (from now on we omit the index
${L^2({\cal M},d\mu_g)}$ in the norms because there is no ambiguity). 
The function 
$x\mapsto ||D_x^\alpha K(T,x,\:.\:|A)||$,
 for $x\in \bar{I}_u$ is continuous from Lebesgue's dominate convergence
 theorem 
since $D_x^\alpha K(T,x,y|A)$ defines a continuous function in $x$ and $y$
and  there is a constant (dependent on $T$ in general) $M_T$ such that
$|D_x^\alpha K(T,x,z|A)|^2 \leq M_T$ for $(x,z)$ which belong in the compact
$\bar{I}_u \times {\cal M}$ and the measure of the manifold is finite. 
The same results holds whenever one keeps fixed $y$ in $I_u$ and integrates
over $x$.
Therefore, let us define
\begin{eqnarray}
P_T^{(\alpha,\beta)} :=
 \left[\sup_{x\in \bar{I}_u} ||D_x^\alpha K(T,x,\:.\:|A)||\right]
 \left[\sup_{y\in \bar{I}_u} ||D_y^\beta K(T,\:.\:, y |A)||\right] \label{P}
\end{eqnarray}
and we have, for any $x,y \in I_u$, the $\lambda_j$-uniform upper bound 
\begin{eqnarray}
| e^{-\lambda_j t }D_x^\alpha \phi_j(x) D_y^\beta \phi^*_j(y)|
\leq P_T^{(\alpha,\beta)}
 e^{-\lambda_j (t-2T)}\:.
\end{eqnarray}
The found inequality proves that the absolute values of the terms of the 
 series 
\begin{eqnarray}
{\sum_{j\in \N}}'
 e^{-\lambda_j t }D_x^\alpha \phi_j(x) D_y^\beta \phi^*_j(y)
\end{eqnarray}
are  
$x,y$-uniformly bounded, for $(x,y,t)\in I_u \times I_u\times
 (2T,+\infty)$,
 by terms of the convergent series (see (30) in \cite{m1})
\begin{eqnarray}
& & {\sum_{j\in\N}}'
 e^{-\lambda_j (t-2T)} P_T^{(\alpha,\beta)} 
=  P_T^{(\alpha,\beta)}\: 
\int_{\cal M}d\mu_g(z) \left\{ K(t-2T,z,z|A) - P_0(z,z|A)\right\}\nonumber\\
 &=& P_T^{(\alpha,\beta)}\: Tr \left\{ K_{(t-2T)} - P_0 \right\} \nonumber\:.
\end{eqnarray}
This holds for any choice of the multindices $\alpha,\beta$ and this entails
 (\ref{pass}), (\ref{ricorda}) and (\ref{ricorda'}).
The final upper bound (\ref{ricorda'}) is a trivial consequence of (99) 
in \cite{m1}
and the fact that the manifold has a finite measure.
 $\Box$\\

\noindent {\bf Sketch of Proof of Theorem 3.1.} Let us fix a coordinate
system in a neighborhood $I_u$ of a point $u\in {\cal M}$, all the following
considerations will be referred to these coordinates, and in particular 
to a couple of points $x,y$ within that neighborhood.
Then, let us consider 
the expression (\ref{Zeta2'}) 
for the $\zeta$ function of the stress tensor.
Employing the eigenvalue equation $A\phi_j = \lambda_j \phi_j$ one can 
rearrange (\ref{Zeta2'})  into
\begin{eqnarray}
 Z_{ab}(s,y|A/\mu^2) &=&
  {\sum_{j\in \N}}'  \frac{2s}{\mu^2}
\left(\frac{\lambda_j}{\mu^2}\right)^{-(s+1)}
T'_{ab}[\phi_j,\phi_j^*, {\bf g}](y) 
 \label{Zeta2+}\:,
\end{eqnarray}
where ($C.C.$ means the complex conjugation of the terms 
already written)
\begin{eqnarray}
T'_{ab}[\phi_j,\phi_j^*, {\bf g}](y) &=& (1-2\xi)\frac{1}{2}
\left(\nabla_a \phi_j(y)\nabla_b \phi_j^*(y) +\phi_j(y)\nabla_a \nabla_b 
\phi_j^*(y) + C.C. \right) \nonumber \\
&+& \left( 2\xi -\frac{1}{2} \right)\frac{g_{ab}(x)}{2}
\left(\nabla_c \phi_j(y)\nabla^c \phi^*_j(y) + \phi_j(y)\Delta \phi^*_j(y)
+ C.C.  \right) \nonumber \\
&+& \frac{1}{2} \left( \frac{g_{ab}(y)}{D} \phi_j(y)\Delta \phi^*_j(y)
- \phi_j(y)\nabla_a \nabla_b
\phi_j^*(y) + C.C. \right) \nonumber\\
&+& \xi \left(  R_{ab}(y) - \frac{g_{ab}(y)}{D}R(y) \right) |\phi_j(y)|^2
- \frac{V'(y) + m^2 }{D} g_{ab}(y) |\phi_j(y)|^2 \nonumber\\ 
&+& \frac{\lambda_j}{D}g_{ab}(y) |\phi_j(y)|^2 \label{f1}\:.
\end{eqnarray}
The stress tensor is then given by (\ref{ztensor}) in {\bf Definition 2.4.}
after the analytic continuation in the variable $s$ of $Z_{ab}(s,y|A/\mu^2)$
given in (\ref{Zeta2+}). Employing {\bf Theorem 2.2} and (\ref{f1}) and 
(\ref{Zeta2+}), we can write down the expression of 
$Z_{ab}(s,y|A/\mu^2)$
 employing also  functions
 $\zeta^{[\alpha,\beta]}(s,y|A/\mu^2)$ defined as in {\bf Definition 2.2}
with the difference that covariant derivatives are employed  instead of
ordinary derivatives.
We get, omitting the arguments $y$ and $A/\mu^2$ in the various $\zeta$
functions for sake of brevity,
\begin{eqnarray}
Z_{ab}(s,y|A/\mu^2)  = \nonumber 
\end{eqnarray}
\begin{eqnarray}
& & (1-2\xi)\frac{s}{\mu^2}
\left( \zeta^{(1_a,1_b)}(s+1) + \zeta^{(1_b,1_a)}(s+1)+
\zeta^{[1_a+1_b,0]}(s+1) + \zeta^{[0,1_a+1_b]}(s+1)
 \right) \nonumber \\
&+& \left(2\xi -\frac{1}{2} \right)\frac{sg_{ab}(y) g^{cd}(y)}{\mu^2}
\left( 2\zeta^{(1_c,1_{d})}(s+1) +\zeta^{[0,1_c+1_{d}]}(s+1) 
+ \zeta^{[1_c+1_{d}, 0]}(s+1) \right) \nonumber \\
&+& \frac{s}{\mu^2} \left[ \frac{g_{ab}(y)g^{cd}(y)}{D} 
\left(\zeta^{[0,1_c+1_{d}]}(s+1) +\zeta^{[1_c+1_{d},0]}(s+1)\right)
- \zeta^{[0,1_a+1_b]}(s+1) \right. \nonumber\\
& & \left. -\zeta^{[1_a+1_b,0]}(s+1)  \right] \nonumber\\
&+& \frac{2s\xi}{\mu^2} \left(  R_{ab}(y) - 
\frac{g_{ab}(y)}{D}R(y) \right) \zeta(s+1)
- \frac{V'(y) + m^2 }{D} \frac{2s g_{ab}(y)}{\mu^2} \zeta(s+1) \nonumber\\
&+& \frac{2sg_{ab}(y)}{D} \zeta(s) \label{f2}\:.
\end{eqnarray}
First of all, we notice that the term proportional to $g_{ab}(y)$ in 
(\ref{fine}) arises from the last term above via item (c) of {\bf Theorem 2.2}
in \cite{m1}.\\
Let us consider the contribution to the stress tensor due to  the terms 
$\zeta^{(1_a,1_b)}(s+1,y|A/\mu^2)$.
Similarly to (101) in \cite{m1}, we can define, for any $\mu_0^2>0$
fixed and $N$ integer $>D/2 + 4$, taking account of {\bf Lemma 2.1} above 
\begin{eqnarray}
\zeta^{(1_a,1_b)}(N,s+1,x,y|A/\mu^{2}, \mu_{0}^{-2}) :=
\frac{\mu^{2s}}{\Gamma(s+1)}
 \int_{0}^{\mu_{0}^{-2}} dt\: t^{s}
 \frac{e^{-\eta\sigma(x,y)/2t}}{(4\pi t)^{D/2}}
t^{N-2} O^{(1_a,1_b)}_{\eta,N}(t;x,y)\nonumber
\end{eqnarray}
\begin{eqnarray}
+ \frac{\mu^{2s+2}}{\Gamma(s+1)}
\int_{\mu_{0}^{-2}}^{+\infty} dt\: t^{s}
\nabla_{(x)a} \nabla_{(y)b}\left[ 
K(t,x,y|A) - P_0(x,y|A) \right] \:.
\end{eqnarray}
Similarly to {\bf Lemma 2.1} in \cite{m1}, one can prove that the function
of $s,x,y$ defined above is continuous in a neighborhood 
 $I\times I_u\times I_u$,
where $I$ is a complex neighborhood of $s=0$ with all of its 
$s$ derivatives, in particular, it is  $s$-analytic therein.
Employing {\bf Lemma 2.1} and 
the item (a) of {\bf Theorem 2.1} one can write also,
for $Re$ $s+1> D/2 +4$,
\begin{eqnarray}
\zeta^{(1_a,1_b)}(s+1,x,y|A/\mu^2) &=& \zeta^{(1_a,1_b)}(N, s+1,x,y|A/
\mu^2,\mu_0^{-2}) \nonumber\\
&-& \left(\frac{\mu}{\mu_0}\right)^{2s+2}\frac{\nabla_{(x)a} 
\nabla_{(x)b}P_0(x,y|A)}{(s+1)\Gamma(s+1)} \nonumber
\end{eqnarray}
\begin{eqnarray}
+ \frac{\mu^{2s+2}}{(4\pi)^{D/2} \Gamma(s+1)}
\sum_{j=0}^N\nabla_{(x)a} \nabla_{(y)b}
 \left( \int_0^{\mu_0^{-2}} dt t^{s-D/2+j} e^{-\sigma/2t} a_{j}(x,y|A) \right)
\label{f4}\:.
\end{eqnarray}
In particular, for $Re$ $s+1> D/2 +4$,
 the left hand side above is continuous in $x,y$ and thus
 we can take the coincidence
limit for $x\rightarrow y$. Noticing that one can also pass the derivatives
under the sign of integration in the right hand side and that 
$\nabla_{(x)a}\nabla_{(y)b} \sigma(x,y)|_{x=y} = -g_{ab}(y)$
and $\nabla_c \sigma(x,y)|_{x=y}=0$, we get
for the right hand side of the expression above multiplied by $2s/\mu^2$
and evaluated for $x=y$
\begin{eqnarray}
 & & 
\frac{2s}{\mu^2}\zeta^{(1_a,1_b)}(N, s+1,y,y|A/\mu^2,\mu_0^{-2}) 
- \frac{2s}{\mu^2}\left(\frac{\mu}{\mu_0}\right)^{2s+2} 
\frac{P_{0ab}(x,y|A)}{(s+1) \Gamma(s+1)}\nonumber\\
&+& \frac{2s}{\mu^2} \frac{\mu^{2s+2}}{(4\pi)^{D/2}\Gamma(s+1)}
\sum_{j=0}^{N} \left\{\frac{a_{jab}(y,y|A) \mu_0^{-2(s-D/2+j+1)}}{s-D/2+j+1}
+ \frac{g_{ab}(y)}{2}\frac{a_{j}(y,y|A) \mu_0^{-2(s-D/2+j)}}{s-D/2+j} \right\}
\nonumber
\end{eqnarray}
where $a_{jab}(y,y|A) := \nabla_{(x)a} \nabla_{(y)b}a_j(x,y|A)|_{x=y}$,
and $P_{jab}(y,y|A) := \nabla_{(x)a} \nabla_{(y)b}
P_0(x,y|A)|_{x=y}$. \nonumber\\
The contribution to the stress tensor,
 namely, to $\frac{d}{ds}|_{s=0}Z_{ab}(s,y|A/\mu^2)/2$ of the
considered term is then obtained by 
continuing the result above as far as
$s=0$, executing the $s$ derivative and multiplying for $(1-2\xi)/2$ the 
final result. Taking account that $\zeta(N,s,x,y|A/\mu^2, \mu_0^{-2})$ 
is smooth in a neighborhood of $s=0$, 
  this lead to, apart from the unessential 
factor $(1-2\xi)$, 
\begin{eqnarray}
\frac{d}{ds}|_{s=0} \frac{2s}{\mu^2}
\zeta^{(1_a,1_b)}(s+1,y|A/\mu^2) \nonumber
\end{eqnarray}
\begin{eqnarray}
 &=& 
\frac{1}{\mu^2}\zeta^{(1_a,1_b)}(N, 1,y,y|A/\mu^2,\mu_0^{-2}) 
- \frac{P_{0ab}(y,y|A)}{\mu^2}\nonumber\\
&+& \frac{1}{(4\pi)^{D/2}}
\sum_{j=0, j\neq D/2-1}^N \frac{a_{jab}(y,y|A)}{\mu_0^{2j-D+2} (j-D/2+1)}
+ \delta_D\left(\gamma +2\ln\frac{\mu}{\mu_0} \right) \frac{a_{(D/2-1)ab}
(y,y|A)}{(4\pi)^{D/2}}\nonumber\\
&+& \frac{g_{ab}(y)}{(4\pi)^{D/2}}
\sum_{j=0, j\neq D/2}^N \frac{a_{j}(y,y|A)}{\mu_0^{2j-D} (j-D/2)}
+ \delta_D g_{ab}(y)\left(\gamma +2\ln\frac{\mu}{\mu_0} \right) 
\frac{a_{D/2} (y,y|A)}{(4\pi)^{D/2}}\:. \label{f5}
\end{eqnarray}
Let us consider the first line in the right hand of (\ref{f2}) side
for a moment. The other terms different from $\zeta^{(1_a,1_b)}(s+1)$
can be undertaken to a procedure similar to that developed above.
The important point is that, once one has performed such a procedure,
 all terms with a factor $g_{ab}(y)$ similar to the terms in the last line
of (\ref{f5}) cancels out each other, and thus, in the final expression
of the first line of the right hand side of (\ref{f1}), no term with
a factor $g_{ab}(y)$ survives. The same fact happens for the second and third
lines of (\ref{f2}).
Since $\zeta^{(1_a,1_b)}(N, 1,y,y|A/\mu^2,\mu_0^{-2}) $ and the derivatives
of heat-kernel coefficients
are continuous in $x,y$ we can compute the right hand side of (\ref{f5})
as a limit of coincidence
\begin{eqnarray}
\frac{d}{ds}|_{s=0} \frac{2s}{\mu^2}
\zeta^{(1_a,1_b)}(s+1,y|A/\mu^2) \nonumber
\end{eqnarray}
\begin{eqnarray}
 &=& 
\lim_{x\rightarrow y}\left\{
\frac{1}{\mu^2}\zeta^{(1_a,1_b)}(N, 1,x,y|A/\mu^2,\mu_0^{-2}) 
- \frac{P_{0ab}(x,y|A)}{\mu^2} \right.\nonumber\\
&+& \frac{1}{(4\pi)^{D/2}}
\sum_{j=0, j\neq D/2-1}^N \frac{a_{jab}(x,y|A)}{\mu_0^{2j-D+2} (j-D/2+1)}
+ \delta_D\left(\gamma +2\ln\frac{\mu}{\mu_0} \right) \frac{a_{(D/2-1)ab}
(x,y|A)}{(4\pi)^{D/2}}\nonumber\\
&+& \left. \frac{g_{ab}(y)}{(4\pi)^{D/2}}
\sum_{j=0, j\neq D/2}^N \frac{a_{j}(x,y|A)}{\mu_0^{2j-D} (j-D/2)}
+ \delta_D g_{ab}(y)\left(\gamma +2\ln\frac{\mu}{\mu_0} \right) 
\frac{a_{D/2} (x,y|A)}{(4\pi)^{D/2}} \right\}\:. \label{f6}
\end{eqnarray}
Moreover, since the function in the limit is continuous,
 the same limit can be computed by identifying the tangent 
space at $x$ with the tangent space at $y$ and thus introducing the 
the bitensor $I_{a}^{a'} = I_{(y)a}^{(x)a'}(y,x)$
of  parallel displacement from $y$ to $x$ as usual.
Employing (\ref{f4}) we finally get
\begin{eqnarray}
(1-2\xi)\frac{d}{ds}|_{s=0} \frac{2s}{\mu^2}
\zeta^{(1_a,1_b)}(s+1,y|A/\mu^2) \nonumber
\end{eqnarray}
\begin{eqnarray}
&=&\lim_{x\rightarrow y}
(1-2\xi) I_{a}^{a'} \nabla_{(x)a'}\nabla_{(y)b}
\left\{ \frac{1}{\mu^2}\zeta(1,x,y|A/\mu^2)
-  H_N(x,y)    \right\}\nonumber\\
&+ & (1-2\xi)  g_{ab}(y) H'(y) \label{f7}
\end{eqnarray}
where, as we said above, the final term proportional to $g_{ab}(y)$
gives no contribution to the final stress tensor because it cancels
against similar terms in the first line of (\ref{f1}).
The explicit form of $H_N$ reads
\begin{eqnarray}
H_N(x,y) &=& \sum_{j=0}^{N} \frac{a_j(x,y|A)}{(4\pi)^{D/2}}
\int_0^{\mu_0^{-2}} t^{j-D/2} e^{-\sigma(x,y)/2t}\nonumber\\
&-& \frac{1}{(4\pi)^{D/2}}
\sum_{j=0, j\neq D/2-1}^N \frac{a_j(x,y|A)}{\mu_0^{2j-D +2}(j-D/2 +1)}
\nonumber\\
&-& \delta_D \frac{a_{D/2-1}(x,y|A)}{(4\pi)^{D/2}}
\left[\gamma + \ln \left(\frac{\mu}{\mu_0} \right)^2 \right]\label{f10}\:.
\end{eqnarray}
The same procedure has to be used for each term in the right hand side of
(\ref{f1}) except for the last term which, as it stands, produces
the last term in the right hand side of (\ref{fine}).
 Summing all contributions,
one gets (\ref{fine}) with $H_N$ in place of $H_{\mu^2}$. 
Anyhow, executing the 
integrations above
  using the results (52) - (58) in \cite{m1} ($D>1)$, expanding 
 $H_N$ in powers and logarithm of $\sigma$
 and taking account of
Comments (1) and (2) after {\bf Theorem 3.1}  above,
we have  that, in the
expansion of $H_N$ one can take account only  of the terms pointed out in 
the item (a) of {\bf Theorem 3.1}; these are the only terms  which
do not contain the arbitrary parameter $\mu^2_0$ (which cannot remain
in the final result). Therefore,  the  part of 
$H_N$ which gives contributions to the final stress tensor coincides  
with $H_{\mu^2}$ given in (\ref{unop}) and (\ref{unop'}).\\
This proves the point (a) of {\bf Theorem 3.1}. The point (b) is trivially 
proven by a direct comparison between (23), (24), (25) in \cite{m1} 
and the 
equation for the coefficients of the Hadamard local solution given in 
Chapter 5 of  \cite{garabedian} which determine completely
the coefficients $u_j$ and $v_j$ of the local solution once  the values of the
coefficients of the leading divergences are fixed for $x \rightarrow y$,
and the coefficients $w_j$ once $w_0$ has been fixed. 
In performing this comparison,
concerning the normalization conditions (\ref{n1}) and 
(\ref{n2}) in particular, notice that the measure used in the integrals
employed in \cite{garabedian} is the Euclidean one $d^nx$ instead of
our measure $\sqrt{g(x)} d^nx$. $\Box$\\

\noindent {\em Acknowledgment.} I am very grateful to D. Luminati
for useful discussions. This work has been financially supported
by a Postdoctoral Research Fellowship of the Department of Mathematics
of the University of Trento.

 \end{document}